\documentstyle[12pt]{article}
\advance\textheight by \topskip
\begin{document}
\begin{flushright}
SU-ITP-95-12\\
hep-th/9506113\\
\today\\
\end{flushright}
\vspace{-0.2cm}
\begin{center}
\baselineskip=16pt

{\Large\bf    DUALITY SYMMETRIC \\
\vskip 0.6 cm
 QUANTIZATION OF SUPERSTRING }\\
\vskip 2cm

{\bf Renata Kallosh} \footnote {  E-mail:
kallosh@physics.stanford.edu}
 \vskip 0.05cm
Physics Department, Stanford University, Stanford   CA 94305\\
\vskip 0.7 cm

\end{center}
\vskip 1.5 cm
\centerline{\bf ABSTRACT}
\begin{quotation}

A general covariant quantization of  superparticle, Green-Schwarz superstring
and a
supermembrane with manifest supersymmetry and duality symmetry is proposed.
This quantization provides a natural quantum mechanical description of curved
BPS-type backgrounds related to the ultra-short supersymmetry multiplets.
Half-size commuting and anticommuting Killing spinors  admitted by such
backgrounds in quantum theory become truncated $\kappa$-symmetry ghosts.  The
symmetry of  Killing spinors under dualities transfers to the symmetry of the
spectrum of states.

GS superstring in the generalized semi-light-cone gauge can be quantized
consistently  in the background of ten-dimensional supersymmetric gravitational
waves.  Upon compactification they become supersymmetric electrically charged
black holes, either massive or massless.  However, the generalized light-cone
gauge breaks S-duality.  We propose a new family of gauges, which we call black
hole gauges. These gauges are suitable for quantization both in flat Minkowski
space and in the black hole background, and they are duality symmetric. As an
example, a manifestly S-duality symmetric black hole gauge is constructed in
terms of the
axion-dilaton-electric-magnetic black hole hair.  We also suggest the U-duality
covariant class of gauges for type II superstrings.

\end{quotation}
\newpage
\baselineskip=16pt
\section{Introduction}
For many years one of the main  goals in quantization of superstrings,
superparticles and supermembranes was to perform the quantization preserving
manifest ten-dimensional supersymmetry and  Poincare invariance. This proved to
be an extremely
complicated problem; its formal solution  involved operations
with an infinite number of ghosts for ghosts. It may happen that eventually we
will find a simple way of  working with this formalism. However, it is not
inconceivable  that the  requirement of ten-dimensional Poincare invariance is
 excessively strong, since it prevents us to study string theory in the
self-consistent gravitational background created by string excitations.

Indeed, the main  idea behind the standard  approach was to use perturbation
theory near the flat  Poincare invariant  background.   However, recently it
was conjectured that among the string eigenstates  there are extreme black
holes. These black holes in some respects behave as ordinary elementary
particles.  A consistent quantization of the string theory should describe such
states as well. Meanwhile, black holes certainly cannot be represented as small
perturbations of the flat Minkowski space.
Some eigenstates of the string theory, which look like extreme black holes in
$d = 4$, can be
considered as gravitational waves in $d = 10$. However, these gravitational
waves are not the usual plane  waves obtained by solving   equations for small
perturbations in the linear approximation (which could be associated with
gravitons), but rather exact solutions of the
full nonlinear gravitational equations.

These solutions have many interesting properties. First of all, the bosonic
configurations have unbroken supersymmetries related to the existence of the
Killing spinors. This leads to  supersymmetric nonrenormalization theorem
\cite{K92} in gravitational theories.
This theorem implies that
all perturbative quantum gravity corrections to the on shell classical action
vanish in
these backgrounds. This is very unusual, since previously the only background
with such a property was the trivial flat Minkowski space. Also, these
solutions, which are called Bogomolny-Prasad- Sommerfield (BPS) states,
saturate the supersymmetric positivity bound
\begin{equation}
M=|Z|\ .
\end{equation}
They have some (nonperturbative) duality symmetries,  which relate to each
other, e.g., electric and magnetic black holes. There are strong indications
that understanding of such states will be essential for  investigation of
nonperturbative properties of string theory.

One may try to find a generally covariant   string quantization procedure which
will preserve  such nonperturbative symmetries as duality invariance. This will
be the main goal which we will try to pursue in the present paper. If
successful, such a program may provide a maximally symmetric quantization
compatible with the non-perturbative structure of the theory. Simultaneously,
it may give us an adequate quantum mechanical description of extreme black
holes, gravitational waves and other BPS states in terms of string theory.

Investigation of the BPS states it the key feature of recent activities in
string theory. However, these states have a dual status. To obtain the
corresponding solutions one is  using   purely classical concepts of space and
time and
classical fields, including the metric of extreme black holes or fundamental
strings and membranes. On the other hand, one is  attributing a certain
quantum-mechanical meaning to these
states.  One would like to get a coherent description of these
``geometries-states'' from the point of view of  string theory.
 As a step towards understanding the quantum-mechanical
meaning of the states describing various geometries, we suggest to find the
place
of the
BPS states in the  quantization of Siegel's $\kappa$-symmetries.

It is known that the gauge $\kappa$-symmetries on the world-line of
the superparticle, on the world-sheet of the Green-Schwarz  superstring and in
the world-volume of the supermembrane share the following important
property:  the spinorial parameter of the gauge transformation has to be
somehow broken into two parts. Only one half of the spinor has to take part
in the transformations, the second part has to be thrown out to comply with
unitarity principle. This implies that the BPS-type geometries with one half of
supersymmetry unbroken may help us to perform the quantization of
$\kappa$-symmetric objects.
In addition, the procedure of quantization via such
geometries may help us to reveal the hidden symmetries of classical
supergravities \cite{CJ} in the spectra of the quantized string.

Another important feature of $\kappa$-symmetries is that for the classical
(not quantized)
superstring  to be invariant under the
$\kappa$-symmetry in  a nontrivial  background requires the background to
satisfy classical equations of motion. Thus even the classical
$\kappa$-symmetric string lives only in configurations
which
solve equations of motions, in particular, in the soliton-type configurations.
We will find that for the quantized string the constraints on the background
are even stronger.

Thus we  propose to implement the  structure of BPS states and the hidden
symmetries of classical  supergravities
\cite{CJ}  into the quantization of the gauge symmetries  of the
superparticle,  of the Green-Schwarz  superstring and
 of the supermembrane. Upon quantization one may expect that
only the S-, T- and U-dualities \cite{SS}--\cite{W} will survive. We
would also like to reveal via quantization the spectrum of the black hole
multiplets with states classified by $USp(4)\times SU(2)$  and the value of the
left-moving charge for the heterotic string,
and by
$USp(8)\times SU(2)$ for type II superstring.

 Our approach will be very closely related to the method of quantization in
the generalized semi-light-cone gauge suggested  some time ago by the
author and by A. Morozov  \cite{K87,KM}. Semi-light-cone gauge is the gauge
where  the two-dimensional metric is in the conformal gauge and the spinor
in the light-cone gauge $\gamma^+ \theta=0$. Generalized semi-light-cone
gauge has a more general constraint on the spinor in terms of a null vector
$n_\mu \gamma^\mu \theta =0, \; n^2 =0$. Also an alternating set of gauges
with two null vectors $n^2 =0, m^2 =0$, $2 mn =1$ was used.

The formalism of ref. \cite{K87,KM} was further developed by
Grisaru, Nishino, and Zanon \cite{Grisaru} and by  Candiello,  Lechner, and
Tonin
\cite{Tonin} for the heterotic sigma model in  a curved background. The most
recent
results are described in \cite{Tonin} where also the reference to previous work
can be
found. Both groups have found the constraints on  possible backgrounds where
the
heterotic string can be quantized consistently in the generalized light-cone
gauge. It
remained unnoticed, however,  that their constraints are satisfied, in
particular,   by
the  background of  extreme electric black holes. We describe the backgrounds
in which the already existing quantization is valid and pay special attention
to the unusual properties of massless black holes.

The generalized light-cone
gauge, in which the heterotic GS string was already quantized, breaks S-duality
since the above mentioned constraints on killing spinors are not satisfied by
magnetic black holes. The basic reason for this is the fact that the Killing
spinors of
electric black holes are constrained by the null condition  whereas in
 magnetic case the constraint is chiral.

In this paper we will introduce a more general class of gauges, which will
allow us to
perform a consistent string quantization without an infinite number of ghosts
in an
arbitrary half-supersymmetric background.  In particular, quantization can be
performed
 in backgrounds including all known types of extreme electric and magnetic
black holes  and   gravitational waves which could be obtained by the black
hole  uplifting to $d = 10$. We will be able to go from  an electric black hole
background to   a magnetic one simply by  changing  a gauge.

Note that we will be able to use these gauges even in Minkowski space, i.e.
even in the
absence of black holes. Thus one can use our method of quantization even in the
limit
when the background becomes trivial. However, this method becomes especially
adequate
for investigation of physical processes in the nontrivial backgrounds
corresponding to
the eigenstates of the string theory, in particular when the background is
given by
massive and/or massless black holes.
The saturation of the supersymmetric positivity bound in the limit
\begin{equation}
M=|Z| \rightarrow 0\
\end{equation}
to the massless states will be included in our analysis.

The paper is organized as follows.  In Sections 1-3 we formulate the general
covariant
quantization scheme for $\kappa$-symmetric theories in arbitrary
half-supersymmetric
backgrounds. In Sections 4-9 we apply the new quantization rules mostly in case
of the
four-dimensional black hole backgrounds.

In Sec. 2 we introduce a new principle of a
general covariant quantization of the $\kappa$-symmetric objects in arbitrary
half-supersymmetric curved backgrounds. We also identify the
background in which
heterotic GS superstring  in generalized light-cone gauge is known to be
quantized consistently, in compactified theory, with the extreme electrically
charged black holes. By passing, we describe the unusual singularities of
massless supersymmetric black holes. In Sec. 3 we formulate the general
duality covariant constraint on the ghosts, which provides a truncation of
infinite
reducibility of
$\kappa$-symmetry. The constraint is given in terms of the  zero mode of the
supercharge of the background. The central charge of the background is used for
algebraic constraint on spinors.  Sec. 4 displays the algebra of
supercharges describing the BPS-states. In Sec. 5 we present duality covariant
gauges for the heterotic string. Sec. 6 is devoted to the procedure of
quantization of $\kappa$-symmetric objects in flat backgrounds, for which the
gauge-fixing condition is defined in the limit of the vanishing background. In
Sec 7. we
describe the path integral for the Green-Schwarz  heterotic superstring in
duality-symmetric gauges. Sec. 8 contains a  detailed description of the
S-duality
covariant class of gauges in terms of the axion-dilaton black hole hair.  In
particular, we
show that the  spinorial part of the black hole gauge behaves as a modular form
of the
weight $\left({1\over 4}, \, -{1\over 4}\right) $ for the left-handed part and
as the
modular form of the weight $\left(-{1\over 4}, \, {1\over 4}\right)$  for the
right-handed
part under S-duality transformations. This is the condition under which the
partition
function on the torus is duality invariant.  In Sec. 9 the U-duality covariant
class of
gauges is described.

In Appendix  A we present some details on supersymmetric gravitational
waves in which the GS string can be quantized consistently in the generalized
light-cone gauge. We also display the four dimensional electrically
charged black holes related to these waves.  In Appendix B  we discuss
 {\it massless} four-dimensional multi-black holes related to ten-dimensional
supersymmetric waves.


\section{\label{General} General covariant quantization of  $\kappa$-symmetry}
  Consider the  Green-Schwarz superstring  (as well as other  objects as
superparticle  and supermembrane which have
local $\kappa$-symmetry)  coupled to a most general on-shell
background\footnote{Under the on-shell background we mean the background
satisfying the classical field equations, which however may have to be
corrected
to
avoid $\kappa$-symmetry anomalies.}
 whose
bosonic part has one-half of unbroken supersymmetries.  This means that the
supersymmetric variation of gravitino\footnote{For configurations solving field
equations and admitting supercovariantly constant spinors defined in
eq. (\ref{grav}) the supersymmetry variations of dilatino and gluino with the
Killing spinor parameter $\epsilon_k$ vanishes, as will be explained later.}
 vanishes,
\begin{equation}
\delta_{susy} \Psi_{\mu} = \hat \nabla _{\mu} \epsilon_k = 0
\label{grav}\end{equation}
for some non-vanishing values of the supersymmetry parameter $\epsilon_k$
which is called Killing spinor. For
the supermembrane
we are looking for the Killing spinors of 11-dimensional supergravity.
For type II
string theory the background gravitino is given by two  $d=10$
Majorana-Weyl spinors
(of opposite chirality for IIA and same chirality for IIB string) and by
one Majorana-Weyl spinor for the heterotic string and for the superparticle. In
all cases the background defines the split of the full spinor
$\epsilon$ into two parts, which we will call Killing spinor
$\epsilon_{k}$ and anti-Killing spinor  $\epsilon_{\bar k}$.
\begin{equation}
\epsilon = \epsilon_{k} +  \epsilon_{\bar k} \ .
\end{equation}
In terms of this split the concept of one-half of unbroken supersymmetry
means that for a given background only the Killing spinor is the non-vanishing
covariantly constant spinor, and its dimension
is equal to one-half of the dimension of the full spinor, or, equivalently, the
dimension of the Killing spinor equals that of the anti-Killing  one:
\begin{equation}
 \hat \nabla _{\mu} \epsilon_{k} = 0\ , \qquad \epsilon_{k} \neq 0 \ ,
\label{kil}\end{equation}
and
\begin{equation}
 \hat \nabla _{\mu} \epsilon_{\bar k} = 0\ , \qquad \epsilon_{\bar k} = 0 \ .
\label{antikil}\end{equation}
In addition,
\begin{equation}
{\rm dim}\; [\epsilon_k]= {\rm dim}\; [\epsilon_{\bar k}] \ .
\end{equation}
Typically, for the opposite sign of electric and magnetic charges of the
background the role of
the Killing and  anti-Killing spinors is reversed:
\begin{equation}
q\rightarrow -q \ , \quad p\rightarrow -p \ ;  \qquad \epsilon_{k} \rightarrow
\epsilon_{\bar k} \ , \quad \epsilon_{\bar k} \rightarrow
\epsilon_{ k} \ ,
\label{anti}\end{equation}
i.e. in the ``anti-background'', characterized by the opposite charges, the
anti-Killing spinor is the  non-vanishing covariantly constant
spinor\footnote{In
presence of $SL(2,Z)$ symmetry, as we will see later,  transformation
(\ref{anti}) is a part of $SL(2,Z)$.}.

We will treat separately the backgrounds
with all supersymmetries unbroken
 (like flat space or Robinson-Bertotti-type
geometries), where
\begin{equation}
 \hat \nabla _{\mu} \epsilon_{k} = 0\ , \qquad \epsilon_{k} \neq 0 \ ,
\label{max1}\end{equation}
and
\begin{equation}
 \hat \nabla _{\mu} \epsilon_{\bar k} = 0\ , \qquad \epsilon_{\bar k} \neq 0 \
{}.
\label{max2}\end{equation}
For the purpose of quantization in flat background or in any other maximally
supersymmetric background (\ref{max1}), (\ref{max2}) the role of the
half-supersymmetric background  (\ref{kil}),
(\ref{antikil})
is to motivate the choice of the gauge-fixing of $\kappa$-symmetric theories
even when the curved background is already absent: there will be still some
special
trace of it in the system, like magnetization in absence of a magnetic field.
In
particular, even in the flat background we will basically use the black hole
hair which
carries the duality property of the theory and represents the property of the
curved space at infinity.

The Killing-anti-Killing split of the full spinor in half-supersymmetric
backgrounds
(\ref{kil}),
(\ref{antikil})
is described  as a specific algebraic relation of the type
\begin{eqnarray}
\epsilon_{k} &=& \chi_{k}  \epsilon \ , \nonumber\\
\nonumber\\
\epsilon_{\bar k} &=& \chi_{\bar k} \epsilon \ ,
\label{general}\end{eqnarray}
where the projectors $\chi_{k}$ and $\chi_{\bar k}$ have the following
properties:
\begin{eqnarray}
\chi_{k} \chi_{\bar k} &=&0 \ , \nonumber\\
 \nonumber\\
\chi_{k} \chi_{k} &=&\chi_{k} \ , \nonumber\\
\nonumber\\
\chi_{\bar k} \chi_{\bar k}&=&\chi_{\bar k} \ .
\label{proj}\end{eqnarray}
The covariantly constant spinors may or may not depend on space-time
coordinates (depending on the configuration and on the frame) but in all cases
the algebraic relation of the type shown above is valid for the constant
part of the spinors which they approach at infinity (for
asymptotically flat space-times).

The basic problem in quantization of  $\kappa$-symmetry
 for the superparticle, for the Green-Schwarz string theory and for the
supermembrane is the following. The gauge symmetry starts with the classical
fields of the  action, but after it is fixed by using the first generation
ghosts, the
ghost  system
also requires a gauge fixing, etc. The origin of the problem is in the fact
that the
generator of the gauge symmetry of the first-generation ghosts in these
theories
is nilpotent on-shell
\cite{K87}.

A  procedure to truncate this infinite set of gauge symmetries was suggested
in  \cite{K87} on the basis of Batalin-Vilkovisky \cite{BV} quantization
method.
We have proposed to use some algebraic constraint on the
$m$th-generation of the ghosts of $\kappa$-symmetry, which makes the
dimension of the truncated ghost equal to one half of the non-truncated one.
The untruncated ghost is a gauge field which requires a next generation of
ghosts, whereas the  truncated  one does not require a gauge fixing, or, to be
more
precise, it  does nor require the next generation of the propagating ghosts.
The truncation was presented in eq. (22) of \cite{K87} in the form
\begin{eqnarray}
&&\sigma^{a}{}_\alpha C ^{\alpha}{}_{(m)} = 0 , \qquad  a=1, \dots , 8, \quad
\alpha
=1, \dots , 16, \nonumber\\
\nonumber\\
&&C^ {\alpha}{}_{(m)} = \tilde \sigma^{\alpha }{}_a  C ^\alpha{}_{(m)}, \qquad
\sigma^{a}{}_\alpha \tilde \sigma^{\alpha }{}_b =0.
\label{oldproj}\end{eqnarray}
The
issues of gauge independence as well as independence on
the truncation procedure were cla\-ri\-fied in this paper. It is clear now that
despite many years of existence of this  formal quantization in terms of
{\it arbitrary} orthogonal projectors $\sigma, \tilde \sigma$ we were lacking
many interesting examples of such projectors which we know now. Moreover,
as we will see later in various examples, duality symmetries are the
symmetries which rotate these projectors, or, in other words, make all of them
possible! The algebraic constraint in our new quantization will come out from
the algebraic constraints which the Killing spinors of the half-supersymmetric
backgrounds satisfy. From this point of view there will be no preference to
any constraint: they will appear on equal footing in the quantized string.

 At the time when the quantization \cite{K87} was performed the set
of algebraic  constraints
which was available was not very rich. In addition to the
standard light-cone condition $\gamma^+ \theta =(\gamma^0 + \gamma^9)
\theta = 0$ we have introduced   a  generalized  light-cone condition, in which
the constraint  on spinors was realized in   terms of two null vectors,
\begin{equation}
n_\mu n^\mu=m_\mu m^\mu = 0 \ ,   \qquad m_\mu n^\mu =  {1\over
2} \ .
\end{equation}
 In
particular, we have imposed the algebraic constraint on the first-generation
ghosts $C_{(1)}$ in the form
\begin{equation}
{\slash \hskip -9pt {m}}
\;{\slash \hskip -6.5pt {n}} \; C_{(1)} =0 \ ,  \qquad   C_{(1)} =  {\slash
\hskip -6.5pt {n}}
\;{\slash \hskip -9pt {m}} C_{(1)}  \ .
\label{trun}\end{equation}
When constraint of this type is imposed, the theory can be quantized as an
irreducible theory with one generation of ghosts of $\kappa$-symmetry. The
gauge, in which the two-dimensional metric was considered in the conformal
gauge and the the fermionic
coordinate of the GS string
$\theta$  in the light-cone gauge, was called the
semi-light cone gauge. When the fermionic variable $\theta$  was constrained in
terms of two-null vectors, as explained above, this gauge was called the
generalized
semi-light-cone gauge.

We have also considered a less restricted differential,
non-algebraic gauge for $\theta$ in which besides the standard
Faddeev-Popov (FP) ghosts  also the propagating Nielsen-Kallosh (NK) ghosts
(related to
Nakanishi-Lathrup fields) had to be taken into account. The role of
these ghosts in BRST quantization was clarified in \cite{BK}. In this gauge
the space-time supersymmetry is realized linearly, as different from the one
with the algebraic constraint.

The partition function for the GS heterotic superstring was constructed in
\cite{K87,KM}. It was   shown to be independent  (at least formally) on the
choice of the truncation
condition on the ghosts and   on the
choice of the gauge condition on the fermionic variable $\theta$.
In particular, in this way one proves  the independence
on the directions
$n, m$ in the choice of constraints. The contribution of the second-class
constraints was taken in the form in which it was derived for the first time
in the series of papers by Fradkin and collaborators \cite{F}.

If one makes a special choice of the vectors
$n_\mu, m_\mu$ one can recover the standard light-cone gauge. This
corresponds to the choice $ {\slash
\hskip -6.5pt {n}} =  \gamma^0 + \gamma^9 $. However, for arbitrary choice
of the vectors there is no need to pick up the direction $9$, it could be any
directions in the nine-dimensional space $1,2,3, 4, 5,6,7,8,9$. Using the
modern language one can summarize this presentations by the statement that
our generalized light-cone gauge has a  T-duality symmetry, $SO(6)$ part
of it, whereas the standard light-cone gauge $(\gamma^0 + \gamma^9)
\theta = 0$ breaks T-duality. A remarkable thing about the proof
\cite{K87,KM} of the independence of the physical states on the choice of the
direction $n,m$ is that it suggests a proof of the T-duality of the states
which arise in the quantization of the string.

Comparing our old truncation condition (\ref{trun}) with
the properties of Killing spinors of the half-supersymmetric backgrounds in
general, given in equations (\ref{general}), (\ref{proj}), we
 see that we have used one particular example of the general projectors.
Our  projectors in (\ref{trun}) obviously satisfy the relations
\begin{equation}
({\slash \hskip -9pt {m}}\;  {\slash \hskip -6.5pt {n}}) ({\slash
\hskip -6.5pt {n}} \;{\slash \hskip -9pt {m}})
=0 \ ,  \qquad
({\slash \hskip -9pt {m}}\;  {\slash \hskip -6.5pt {n}})
^2 = ({\slash \hskip -9pt {m}}\;  {\slash \hskip -6.5pt {n}}) \ , \qquad
({\slash \hskip -6.5pt {n}}\;  {\slash \hskip -9pt {m}})^2 =
({\slash \hskip -6.5pt
{n}}\;  {\slash \hskip -9pt {m}}) \ .
\end{equation}
Therefore, the  generalized $(m,n)$ light-cone-type truncation of fermionic
symmetry, which was  used in
\cite{K87,KM}, is  associated with the Killing
 spinors of the backgrounds for which
\begin{equation}
\chi_k (g. l.c.)= {\slash \hskip -6.5pt {n}}\;  {\slash \hskip -9pt
{m}} \ ,
\qquad
\chi_{\bar k} (g. l.c.)= {\slash \hskip -9pt {m}}  {\slash \hskip
-6.5pt {n}} \ .
\end{equation}
The heterotic GS superstring $\sigma$-model was constructed in \cite{Grisaru}.
It was discovered there that the quantization of the heterotic
string  in generalized $(m,n)$ light-cone gauge is consistent only when the
background is constrained in a specific way, the constraint being stronger
than the requirement that the background satisfies classical equations of
motion. The most recent quantization of the GS heterotic
$\sigma$-model was performed in
\cite{Tonin}.  Both groups have studied the issues of $\kappa$-anomalies.

We would like to reformulate here the constraint on the background as given
in \cite{Grisaru,Tonin} in a form which is suitable for the generalization to
most general
BPS-states. These states correspond to  backgrounds of the superstring,
 which have fermionic isometries related to Killing spinors of dimension equal
to
one-half of the full spinor.
To classify those
isometries we will introduce the following definitions.
\begin{itemize}
\item  {\it Supersymmetric gravitational waves} are the supergeometries
whose  bosonic part  admits a supercovariant Killing spinor and  a null Killing
vector.
\item  {\it Supersymmetric gravitational waves of electric type }  are the
supergeometries  whose  bosonic part  admits a supercovariant Killing spinor
 satisfying the null constraint ${\slash \hskip -6.5pt {n}}
\epsilon_k =0$ where $n$ is a null Killing vector.

\item  {\it Supersymmetric pp-waves } are the special set of
Supersymmetric gravitational waves of electric type whose bosonic part
admits a  covariantly constant null Killing vector.

\item  {\it Supersymmetric gravitational waves of magnetic type }  are the
supergeometries  whose  bosonic part  admits a supercovariant Killing spinor
 satisfying a chiral  constraint $(1-\Gamma^5)\epsilon_k=0$,
 where $1-\Gamma^5$ is a chiral projector in any $SO(4)$ subspace of the full
$SO(1,9)$ tangent space of the supergeometry. They also admit at least one null
Killing vector.

\item  {\it Supersymmetric gravitational waves of
electro-magnetic  type } are the supergeometries whose bosonic part
admits at least one null Killing vector and the supercovariant Killing spinor
satisfies the constraint which is neither null nor chiral.

\end{itemize}

 The constraints on the backgrounds in which the heterotic string can
be
quantized consistently in the generalized light-cone gauge were presented in
\cite{Grisaru,Tonin}. It remained
unnoticed
that these constraints require the background to correspond to electric BPS
states.
In more precise form our analysis shows the following.

{\it The  heterotic GS string can be consistently quantized in $(m,n)$
light-cone
gauge
 in the background of ten-dimensional  supersymmetric gravitational
waves of electric type
 or in any  compactified  form of it. In particular when the
supersymmetric wave is reduced to four-dimensional theory, one gets the
 most general electrically charged extreme black-hole-type
solutions of heterotic string}.

Indeed, the background has to admit algebraically constrained
 covariantly
constant spinors to comply with the requirements of truncation of gauge
symmetry. When the algebraic constraint on the ghost
is ${\slash \hskip -6.5pt {n}}
C_{(1)} =0$ , we are looking for the most general
configurations  which admit covariantly constant spinors satisfying this
constraint. Since the corresponding vector $n_{\mu}$ is null,
we are simultaneously looking for geometries which admit a null Killing
vector. Indeed the constant null vector of the flat background
will become covariantly constant in the curved
background. This brings us to the backgrounds which have one covariantly
constant null vector: to supersymmetric pp-waves \cite{BKO1} which we have
called SSW. The metric is that of Brinkmann, and other fields are adjusted for
supersymmetry. The configuration may depend on $u, x^1, \dots x^8$ but has to
be
independent on
$v$. If both null vectors are used in the alternative type gauges one can relax
covariant constancy of both vectors and look only for the
supersymmetric backgrounds which admit two null Killing vectors. Those
depend only on transverse coordinates $x^1, \dots x^8$ but have to be
independent on both light-cone coordinates $u,v$. These configurations
include fundamental strings \cite{Da1}, generalized fundamental strings
\cite{BEK} etc. These configurations also fall into the definitions of
gravitational waves, since they admit a null Killing vector. The chiral null
models of Horowitz and Tseytlin \cite{HT} also admit a supercovariantly
constant
Killing spinor satisfying the null constraint and
 belong to the class of gravitational waves of electric type.   We present the
relevant details on these configurations and their relation to the most general
supersymmetric electrically charged multi-black-hole-type solutions  in the
Appendices A and B. As an example  we describe here
the  spherically symmetric electrically charged black holes in which the
heterotic string
is
quantized consistently. They have the
following four-dimensional canonical metric:
\begin{equation}
ds^2 = e^{2\phi} g^{-2}  dt^2 -  e^{-2\phi} g^2 (d\vec x)^2 \ ,
\label{elbh}\end{equation}
where the four-dimensional dilaton field is given by
\begin{equation}
e^{-2\phi}= {1\over g^2} \left( 1 +  {4 m G_N  \over  r} + { 4
              g^2 (N_L -1)\over r^2}
\right)^{1\over 2} \ .
\label{dilaton}\end{equation}
In eq. (\ref{elbh}) $G_N$ is the Newton constant and $N_L$ is the
contribution from the left moving oscillators.
The Bogomolny bound in notation of \cite{Sen95}  (for
$G_N=2$) states that the mass of the black hole is equal to the central charge
of the graviton multiplet,
which in turn is defined by the right-moving electric charge as well as by the
combination of the left-moving charge and $N_L$:
\begin{equation}
m^2 = |Z|^2 = {\vec Q_R \over 8 g^2} = {g^2\over 8} \left({\vec Q_L^2\over g^4}
+ 2
N_L -2\right) \ .
\end{equation}
 The relation of this  solution to
 supersymmetric ten-dimensional gravitational waves
 and to four-dimensional black holes   \cite{Sen95} is explained in Appendix A.
The black hole solution (\ref{elbh})
interpolates
nicely between the $a=1$ and $a=\sqrt 3$ heterotic string supersymmetric
electrically charged black
holes. Indeed,  for
$a=1$ solutions the left moving charge $Q_L$ is vanishing and therefore   the
dilaton is given by the  harmonic function,
\begin{equation}
e^{-2\phi}= {1\over g^2} \left( 1 +    {8 m \over
r}                                                 + { 16 m^2\over r^2}
\right)^{1\over 2}= {1\over g^2} \left (1+ {4 m \over
r}\right ) \ .
\end{equation}
For $a=\sqrt 3$ we have $N_L=1$. For
this solution the dilaton is given by the square root of the harmonic function,
\begin{equation}
e^{-2\phi}= {1\over g^2} \left( 1 +  {8 m   \over  r} \right)^{1\over 2} \ .
\end{equation}
Note, however that the general  solution (\ref{elbh}), (\ref{dilaton})
corresponds not  to black holes with arbitrary dilaton coupling $a$,
 but to more generic dimensionally reduced supersymmetric
gravitational waves. It has been noticed by Behrndt \cite{Klaus2}
that there exists a massless black hole configuration  in (\ref{elbh}),
(\ref{dilaton}). Indeed, the two-parameter solution with $m=|Z|=0$ and
$N_L=0$ has the form (\ref{elbh}), where the canonical four-dimensional
metric is:
\begin{equation}
ds^2 =  \left( 1 -  { 4
              g^2 \over r^2} \right)^{-{1\over 2}}  dt^2
 -   \left( 1 -  { 4
              g^2 \over r^2}
\right)^{1\over 2}  d\vec x^2 \ ,
\label{dilatonmassless}\end{equation}
and the four-dimensional dilaton is
\begin{equation}
e^{-2\phi}= {1\over g^2}  \left(   {r^2 -  4
              g^2 \over r^2}
\right)^{1\over 2} \ , \qquad     e^{-2\phi}  (r\rightarrow    \infty)
 \equiv   {1\over g^2} \ .
\label{dilatonmassless2}\end{equation}
One of the striking properties of massless dilaton black
holes is the appearance of a new type of singularity.  Massive
extreme black holes have the singularity and the horizon both situated at
$r=0$ ( the only non-singular solution is the pure magnetic
$a=1$ massive extreme black hole in stringy frame).
Massless states are getting  an additional singularity at $r \not = 0$. The
position of the singularity is related to the string coupling constant.

The electric solution is singular at $r = 2g$ (and at $r=0$). At $r \rightarrow
2g$ the dilaton blows up! As different from massive electrically charged black
holes, which near singularity $r=0$  have small gauge coupling, the massless
electrically charged black holes have infinite coupling near the singularity $r
= 2g$. The singularity at $r=0$ and the fact that the string coupling becomes
small are irrelevant for massless electric black holes,
\begin{equation}
(e^{2\phi})_{r\rightarrow  2g}^{\rm el} \rightarrow \infty \ .
\label{dilcoupl}\end{equation}

The magnetic massless solutions
\footnote{These solutions do not fit into the
dimensionally reduced supersymmetric gravitational waves of
electric type, for which the quantization performed in \cite{Grisaru,Tonin}
can be applied directly. However, it will be shown later that for
supersymmetric waves of magnetic type there exists a more general gauge
condition, in which the quantization can be performed.} has the form
\begin{equation}
ds^2 =  \left( 1 -  { 4
               \over  g^2 r^2} \right)^{-{1\over 2}}  dt^2
 -   \left( 1 -  { 4
               \over g^2 r^2}
\right)^{1\over 2}  d\vec x^2 \ .
\label{dilatonmasslessmagn}\end{equation}
and the four-dimensional dilaton of the magnetic solution is:
\begin{equation}
e^{2\phi}= g^2  \left( 1 -  { 4
               \over g^2  r^2}
\right)^{-{1\over 2}} \ , \qquad     e^{2\phi}  (r\rightarrow    \infty)
\equiv   g^2 \ .
\label{dilatonmassless3}\end{equation}
This solution is singular at $r = {2\over g} $ (and at $r=0$). Here again we
have the picture quite opposite to the usual properties of massive magnetically
charged dilaton black holes. Near the singularity $r = {2\over g} $ the string
coupling vanishes! Indeed, in this case
\begin{equation}
(e^{2\phi})_{r\rightarrow  2g}^{\rm magn} \rightarrow 0 \ .
\label{dilmag}\end{equation}
It is particularly important that even in the limit of the vanishing ADM mass
of the black
hole  considered above the configuration still has unbroken supersymmetry and
the Killing spinor
satisfies the same constraint as for the black holes with the non-vanishing
mass.

The multi-black-hole solutions generalizing those in eq.
(\ref{elbh}) as well as the most general known to us stationary supersymmetric
solutions can be found in the
Appendix A. The massless black holes, some of their properties, including
singularities, as well as more general massless black hole and multi-black hole
solutions are presented in Appendix B.

All those configurations have one-half of unbroken supersymmetry and
therefore  the heterotic string can be quantized consistently in these
backgrounds.

\section{New truncation of $\kappa$-symmetry}

The backgrounds with half of
supersymmetries unbroken, which were intensively studied in the recent
years, see e.g.
\cite{DUFF,Tsey} for a review,
offer a much more general class of truncation of $\kappa$-symmetries. To be
more explicit, we may use any  BPS background to get the orthogonal projectors
needed for truncation of $\kappa$-symmetry and defined in eqs.
(\ref{oldproj}) in old form and in eq. (\ref{proj}) in a form related to
 the Killing-anti-Killing split in eqs. (\ref{kil}), (\ref{antikil}) and
(\ref{general}). Moreover, all solutions of the Killing equations
(\ref{grav}) for eleven- and  ten-dimensional supergravities which are not
known yet and still wait to be discovered, are already included in
quantization.

{\it  Consistent quantization of truncated
$\kappa$-symmetry is possible in the backgrounds with one half of unbroken
supersymmetry.
The  integrability condition for the
existence of  Killing spinors of the bosonic part of the background is the
consistency condition for the quantization.
This defines
the curved
superspace
in which quantized
$\kappa$-symmetric objects exist}.

The reason why the geometries with
one-half
of supersymmetries unbroken (BPS-states) are singled out is related to
the fact that the dimension of the truncated $\kappa$-symmetry ghost has to be
one
half
of the  untruncated one to preserve unitarity of the quantization and the
correct
counting of the physical degrees of freedom.

The most general truncation of $\kappa$-symmetry can be achieved in terms
of the most general algebraic constraint, which the Killing spinors of the
half-supersymmetric backgrounds satisfy. Our goal is not to use any specific
background for this purpose, but the most general one which may define the
Killing-anti-Killing split of the spinor. The key role in our quantization of
the
$\kappa$-symmetric systems  belongs to the supercharge of the background.

The supercharge of the gravitational supersymmetric theory was defined
by Teitelboim \cite{Teit}  in asymptotically flat spaces as the surface
integral
in terms of the gravitino $\Psi_{\mu}$ field of the configuration, solving the
field equations:
\vskip -.3cm
\begin{equation}
{\cal Q} = \oint_{\partial \Sigma}  d\Sigma_{\mu\nu} \gamma^{\mu\nu\lambda}
\Psi_\lambda  \ .
\end{equation}
\vskip -.3cm
The surface over which the integration has to be performed depends on the
choice of configuration. In all cases it is the same surface the integration
over
which defines the ADM mass of a given system or the  ADM mass per unit  area
(length).
 The on-shell backgrounds with one-half of supersymmetry
unbroken in bosonic sectors have the vanishing supersymmetry variation of the
gravitino, when the parameters are Killing spinors, as defined  in eqs.
(\ref{grav}), (\ref{kil}), (\ref{antikil}).
\vskip -.3cm
\begin{equation}
{\cal Q}_{k}  =  \oint_{\partial \Sigma}   d\Sigma_{\mu\nu}
\gamma^{\mu\nu\lambda}
\hat \nabla _\lambda \epsilon_k =0 \ .
\label{killingcharge}\end{equation}

Let us stress that $\kappa$-symmetry of the classical action is preserved
only in the on-shell superbackground. This means that the bosonic part of the
background in absence of fermions in the solutions has to solve classical
equations of motion. For the heterotic string this means that the vanishing of
supersymmetry variations of gravitino $ \Psi_\mu$ is  a sufficient condition
for the
vanishing of the  supersymmetry variations of dilatino
$\lambda$ and gluino
$\chi$. To prove it one can
use the Nester construction in the form used in  \cite{HL}. Thus we
start with
\begin{equation}
\delta_{\epsilon_k} \Psi_\mu =0 \hskip 2 cm \Longrightarrow \hskip 2 cm
N^{\mu\nu}
=\bar\epsilon_k\, \gamma^{\mu\nu\lambda}
\hat \nabla _\lambda \epsilon_k =0 \ .
\end{equation}
Using eq. (3.14)  from \cite{HL}  we get
 \begin{equation}
\hat \nabla _\mu N^{\mu 0} = ( {\delta_{\epsilon_k}
\lambda})^\dagger ( \delta_{\epsilon_k}
\lambda)  +  ( {\delta_{\epsilon_k}
\chi} )^\dagger ( \delta_{\epsilon_k}
\chi ) + {\rm field \; eqs.} =0 \ .
\end{equation}
Since the field equations have to be satisfied for the background of the
superstring even before quantization, we conclude that the existence of a
supercovariantly constant spinor (\ref{killingcharge}) is necessary and
sufficient condition for the bosonic background to have  half of
supersymmetries unbroken. The full background  corresponding to such bosonic
backgrounds  has fermionic isometries of dimension equal to one half of the
dimension of
the full fermionic part of the superspace.

 For anti-Killing spinors the supercharge is not vanishing. For the black hole
multiplets it defines the
so-called superhair of the black hole:
\vskip -.3cm
\begin{equation}
{\cal S}_{\rm superhair} \equiv {\cal Q}_{\bar k}  =  \oint_{\partial \Sigma}
d\Sigma_{\mu\nu}
\gamma^{\mu\nu\lambda}
\hat \nabla _\lambda \epsilon_{\bar k}  \ .
\end{equation}
\vskip -.3cm
\noindent The concept  of the {\it superhair}  was defined for the first time
for
 extreme Reissner-Nordstr\"om  black holes in  \cite{Aich} and studied more
recently in the context of more general extreme black holes in \cite{BrKO}.

We postulate the  new truncation of infinite reducibility of $\kappa$-symmetry
by requiring  some odd (even) generation $\kappa$-symmetry ghost to be a
commuting  $m=2n+1$ (anticommuting  $ m=2n$)
 fermionic zero mode of the {\it zero supercharge condition} \footnote{The
fermionic coordinate of the string  $\theta$ is included in this
set
 as $p=0$ case.}
\vskip -.3cm
\begin{equation}
{\cal S}_{\rm ghost} =  \oint_{\partial \Sigma}   d\Sigma_{\mu\nu}
\gamma^{\mu\nu\lambda}
\hat \nabla_\lambda  C_{(m)} =0 \ .
\label{zerocharge}\end{equation}

In  other words, we require the parameter of the $\kappa$-symmetry
transformation to be a Killing spinor of the geometries associated
with the  states which saturate the supersymmetric positivity bound and the BPS
bound. This gives a
perfect and universal accomplishment of the goal: to truncate the infinite
reducibility of any $\kappa$-symmetric theory.
 The integration in (\ref{zerocharge}) has to be
performed over the suitably defined surface.

For example, in the context of the
ten-dimensional heterotic string toroidally compactified to four dimensions,
the relevant surface defining the ADM mass of the  four-dimensional black
holes is the two-sphere at spatial infinity times the internal space
$\partial \Sigma = S^2_\infty \times T^6 $. However,
for each particular class of problems the choice of a surface in the definition
of the  supercharge depends on the class of configurations which
are interesting in  specific problems. For example, supersymmetric
fundamental string  \cite{Da1} provides a supercharge per unit of length, and
the surface  $\partial \Sigma$
is an eight-dimensional space-like surface.
One may expect that for the
supermembrane the surface will be the same as the one defining the
ten-dimensional ADM mass of the black holes. Various surfaces
for the quantization of $\kappa$-symmetric objects in various phases still
have to be identified. In particular, for  various supersymmetric $p$-branes
and black holes in diverse dimensions there will be all kind of relevant
surfaces. A non-trivial situation may occur when the mass of the black holes
vanishes due to the shrinking of the corresponding area, as shown by Strominger
for type II B string theory \cite{Strom}. However even in this limit the
background still provides  a Killing spinor
suitable for quantization.

The definition of the
supercharge (as well as the definition of the ADM mass) does not violate
general covariance, it is just the way to describe the gravitating systems
with special behavior at infinity.

Upon integration the truncation condition acquires a form of the algebraic
constraint on the Killing spinor in the half-supersymmetric bosonic background
of  the form
\begin{equation}
{\cal S}_{\rm ghost} = \chi_{\bar k}  C_{(m)} =0 \ , \qquad C_{(m)} = \chi_{ k}
C_{(m)} \ .
\end{equation}
The wonderful property of the half-supersymmetric backgrounds is that they
admit both commuting and anticommuting Killing spinors.Therefore we may
use this algebraic condition either on the anticommuting variables $\theta$ in
the
classical action corresponding to the unitary gauge with non-propagating
$\kappa$-symmetry ghosts, or on any generation of the commuting-anticommuting
ghosts, since their statistics alternates.

Thus we propose to truncate the infinite reducibility of $\kappa$-symmetry
identifying the fermionic ghosts with the asymptotic value of the Killing
spinor of the bosonic part of the background.
The constraint (\ref{zerocharge}) is the most general constraint which allows
the truncation and consistent quantization in a given background.
Simultaneously
it restricts the backgrounds by requiring them to admit supercovariantly
constant spinors of the dimension ${1\over 2}$  of the dimension of the
original spinor.

Thus  what remains is to find  the most
general background for each theory (superparticle in arbitrary
dimensions, GS type II superstring,  the heterotic string, and the
supermembrane) which has one half of unbroken supersymmetry. This would
supply us with the most general algebraic constraint for the truncation of
$\kappa$-symmetry in each of the above mentioned theories.

The best known to us example of such kind is the list of all metrics
admitting supercovariantly constant spinors in $N=2$ supergravity
(and more recently in $N=4$), performed by Tod \cite{Tod}. In $N=2$
supergravity interacting with $N=2$ matter  Tod has listed all metrics and
has found all supercovariantly constant spinors. The reason for this was
partially related to the fact that $N=2$ supersymmetry with two Majorana
spinors, or one Dirac spinor is suitable for the use of the
highly developed Newman-Penrose formalism with commuting Dirac
spinors.

In most of the other cases related to $N=4$ supergravity with matter, or in $
N=8$ supergravity in compactified theories, or directly in
$4\geq d\geq 11$ there is a rapidly growing amount of information about
bosonic configurations with one-half of unbroken supersymmetries. Those
configurations are related by dualities,  by dimensional reduction, and/or
uplifting. Examples include extreme black holes, fundamental strings,
 p-branes,
pp-waves, dual strings, and dual waves. However, at present we do not have all
solutions of
integrability conditions for the existence of Killing spinors in higher
dimensional supersymmetries.


\section {Extreme black hole superspace and the supercharge algebra}

We are interested in  the  quantization of  the $\kappa$-symmetric
 superstring the  background superspace with
the following properties: it is an on-shell superspace (in the first
approximation prior to the the study of $\kappa$-symmetry anomalies). However,
only special on-shell superspaces are allowed: half of fermionic directions are
isometries. This means that the system of coordinates exists in which the
configuration is independent on half of fermionic coordinates of the
superspace.
This corresponds to the fact that
the bosonic part of the background
admits supercovariantly constant Killing spinors. The
 supercharge which forms the Clifford algebra, defining the ultra-short
supermultiplet
of  string excitations is build in terms of anti-Killing spinor whose
dimension is the same as that of the Killing spinor.  Such backgrounds  allow
the
general covariant truncation of infinite reducible $\kappa$-symmetry and
consistent quantization of the superstring.

 Most of our
attention here will be directed to the extreme four-dimensional black hole
supermultiplets and their spectra. Therefore we will describe the algebra of
the
supercharges
${\cal Q}
$ representing the ten-dimensional Majorana-Weyl or Majorana spinors as
$d=4, N=4$  or $d=4, N=8$ spinors.  However, the strategy for quantization of
generic $\kappa$-symmetric objects remains the same if one is interested in the
spectrum of higher-dimensional configurations.

The algebra which the supercharges ${\cal Q} $ satisfy in backgrounds with half
of supersymmetries unbroken is most conveniently described for massive states
at
rest  with $M= |Z|$
in terms of a $2N$-component spinors \cite{FSZ} \footnote{Notation of this work
are used in this section.}.
In doublet form they are given by
\begin{equation}
 Q_{\alpha}^{ a} = \left (\matrix{
Q_{\alpha}^{ i} \cr
\cr
 Q^{* \alpha   i}\cr
}\right ), \quad Q_{\alpha}^{ a} = Q_{\alpha}^{ i} \  \  \mbox{for} \ \
a=1,\dots ,  N ,
\quad Q_{\alpha}^{ a}= Q^{* \alpha   i} =\epsilon^ {\alpha \beta} Q^{* i}_
\beta \  \ \mbox{for} \ \
a= N+1,\dots , 2N.
\end{equation}
These spinors satisfy a symplectic reality condition
\begin{equation}
Q_{\alpha}^{*a} = \epsilon ^{\alpha\beta} \Omega_{ab} Q_{\beta}^{ b}
\end{equation}
with
\begin{eqnarray}
\Omega^{ab} = - \Omega_{ab}  = \pmatrix{
0&  {\rm I}\cr
-   {\rm I} & 0 \cr } \ , \end{eqnarray}
\begin{eqnarray}
\{  Q_{\alpha}^a ,  Q_{\beta }^b \} = \epsilon _{\alpha \beta} \pmatrix{
Z  & |Z|  {\rm I}\cr
\cr
- |Z|  {\rm I} & Z^{*} \cr } \equiv  \epsilon _{\alpha \beta} \; {\cal Z} ^{ab}
\label{Z} \ .
\end{eqnarray}
The $2N\times 2N$ matrix ${\cal Z} ^{ab}$ is written in terms of $N \times N$
numerical antisymmetric complex matrix $Z^{ij}$ and $| Z| \delta^{ij} $. The
numbers $Z^{ij}$ are the eigenvalues of the central charge operators in a given
supermultiplet. For the purpose of quantization we need to consider the special
BPS case when
\begin{equation}
- Z Z^{* } =Z Z^\dagger = \delta^ {il}  |Z|^2    \ .
\end{equation}
If we would have a massive multiplet without central charges the algebra
generating the states would be
\begin{eqnarray}
\{  Q_{\alpha}^a ,  Q_{\beta }^b \} = \epsilon _{\alpha \beta} \pmatrix{ 0
 & M \cr
- M   & 0 \cr } \equiv  \epsilon _{\alpha \beta} \; \Omega^{ab} M \ .
\label{massive}\end{eqnarray}
The physical massive states of $N$-extended supersymmetry without central
charges are classified according to $USp(2N)\times SU(2)$.
When  the extreme constraint
  is relaxed, i.e. $M \neq |Z|$, the algebra is
\begin{eqnarray}
\{  Q_{\alpha}^a ,  Q_{\beta }^b \} =   \epsilon _{\alpha \beta} \pmatrix{
Z & M {\rm I}\cr
- M  {\rm I} & Z^* \cr } \equiv  \epsilon _{\alpha \beta} \; {\cal P} ^{ab} \ .
\end{eqnarray}
The $USp(2N)$ symmetry of eq. (\ref{massive}) is broken in presence of
central charges $Z^{ij}$.  The effect of $Z^{ij}$ is to reduce the $U(N)$
symmetry
of supersymmetry algebra to subgroups of $U(N)$ which leave the complex
skew-symmetric numerical matrix $Z^{ij}$ invariant. Therefore one has to
classify
the orbits of the two-fold antisymmetric representation $[N]_2$ of $U(N)$. From
all possibilities to have  central charges for the purpose of quantization we
are
interested only in one: in {\it special critical  orbit with $USp(N)$
invariance}.
In this case only the central charge matrix ${\cal Z} ^{ab}$ has the
properties
required to fix $\kappa$-symmetry: the rank of this matrix\footnote{The rank of
the matrix is the maximal size of its invertible square minor.} equals $N$
whereas the dimension is $2N$.

Thus we would like to use
only the backgrounds with the supercharge satisfying the algebra (\ref{Z}).
The condition of exactly one half of unbroken supersymmetry requires all
non-vanishing eigenvalues of
$||Z||$
coincide. Under this condition exactly half of generators ${\cal Q}_{k}$ drop
from
the algebra, and $USp(2N)$ symmetry is broken down to $USp(N)$. Let us stress
that the extremality condition $M=|Z|$ means that in the absence of central
charges,
there are no massive states. This is actually well known from black hole
theory:
supersymmetric extreme black holes require the presence of central charges
\cite{Aich,US}.

Any complex antisymmetric matrix $Z^{ij}$ can be brought to the normal form
using some $U(N)$ rotation.  For example,  for $N=4$ and for $N=8$ respectively
we get
\begin{equation}
\tilde Z_{ij} = i \sigma_2 \left (\matrix{
|Z| & 0  \cr
0 & |Z|  \cr
}\right ) \ , \qquad
\tilde Z_{ij} = i \sigma_2
\left (\matrix{
|Z| & 0 & 0 & 0 \cr
0 & |Z| & 0 & 0 \cr
0 & 0 & |Z| & 0 \cr
0 & 0 & 0 & |Z| \cr
}\right ) \ ,
\end{equation}
where $|Z|$ is a  non-negative real number.
When the central charge matrix is in the normal form, one can perform a
symplectic
transformation over $Q_\alpha^a \rightarrow  T^a{}_b Q_\alpha^a = S_\alpha^a$
with
some numerical matrix T \cite{FSZ}. As a result, we get for supersymmetry
generators in the   ``electric black hole''  basis:
\begin{equation}
\{  S_{\alpha}^a ,  S_{\beta }^b \} =  \epsilon _{\alpha \beta} \;
 \tilde {\cal Z} ^{ab}
 = \pmatrix{
0 & |Z|(1+ \sigma_3) \cr
-|Z|(1+ \sigma_3) & 0 \cr} \ .
\label{centr}\end{equation}
Now it is easy
 to define the projectors which separate the vanishing supercharge ${\cal
Q}_{\bar k}$ from the anti-Killing one ${\cal Q}_{k}$:
\begin{equation}
{\cal Q}_{k} = {1\over \sqrt 2} (1- (-1)^{a+1} )   S_{\alpha}^a \ ,  \qquad
{\cal Q}_{\bar k} = {1\over \sqrt 2} (1 + (-1)^{a+1} )   S_{\alpha}^a \ .
\end{equation}
This means that in our basis the even in $a$ components of the symplectic
spinor $S_{\alpha}^a$ commute, whereas the odd components generate the
spectrum.
We introduce the notation
\begin{eqnarray}
({\cal Q}_{k })_{\alpha}^{m} &=& {1\over \sqrt 2} (1- (-1)^{a+1} )
S_{\alpha}^a ,  \qquad m\equiv  2a= 2,4, \dots  , 2N \ , \\
\nonumber\\
({\cal Q}_{\bar k})_{\alpha}^{\hat m} &=& {1\over \sqrt 2} (1- (-1)^{a+1} )
 S_{\alpha}^a \ ,  \qquad \hat m \equiv  2a+1 = 1,3, \dots , 2N-1 \ .
\end{eqnarray}
Each of the   $N$-component  Killing and anti-Killing spinors
$ ({\cal Q}_{k })_{\alpha}^{m} $  and
$ ({\cal Q}_{\bar k })_{\alpha}^{\hat m} $ satisfy the symplectic reality
condition which does not mix them:
\begin{equation}
({\cal Q}_{k })_{ \alpha}^{* m} = \epsilon ^{\alpha\beta} \Omega_{mn} ({\cal
Q}_{k })_{\beta }^{n}  \ , \qquad ({\cal Q}_{\bar k })_{ \alpha}^{* \hat m} =
\epsilon ^{\alpha\beta} \Omega_{\hat m \hat n} ({\cal Q}_{k })_{\beta }^{\hat
n} \ .
\end{equation}
In doublet form they are given by the $N$-component spinors
\begin{equation}
({\cal Q}_{ k})_{\alpha}^{ m} = \left (\matrix{
({\cal Q}_{ k})_{\alpha}^{ p} \cr
\cr
({\cal Q}_{ k})^{* \alpha   p}\cr
}\right ) \ ,
\qquad
({\cal Q}_{\bar k})_{\alpha}^{\hat m} = \left (\matrix{
({\cal Q}_{\bar k})_{\alpha}^{\hat p} \cr
\cr
({\cal Q}_{\bar k})^{* \alpha  \hat p}\cr
}\right ) \ , \qquad p, \hat p = 1, \dots , {N\over 2} \ .
\end{equation}
In normal basis we  may rewrite the supercharge algebra (\ref{Z})  as follows:
\begin{eqnarray}
\{ ({\cal Q}_{k })_{\alpha}^{m} ,  ({\cal Q}_{k })_{\beta}^{n} \} &=& 0 \ , \\
\nonumber\\
\{  ({\cal Q}_{\bar k})_{\alpha}^{\hat m}  , ( {\cal Q}_{\bar k})_{\beta }
^{\hat n}  \} &=&   |Z| \;
\epsilon _{\alpha \beta} \Omega^{\hat  m  \hat n} \ , \\
\nonumber\\
\{ ({\cal Q}_{k })_{\alpha}^{m} ,  ({\cal Q}_{\bar k })_{\beta}^{\hat n} \} &=&
0 \ .
\label{normal}\end{eqnarray}
Thus, our extreme black hole basis for extended supersymmetry in a normal form
represents two $USp(N)$ doublets instead of the one original $USp(2N)$ doublet
$Q_{\alpha}^a $. Only one of those doublets (anti-Killing one) forms the
Clifford algebra, the second one (Killing one) anticommutes with both of
them.  The Clifford vacuum $\Omega$ is annihilated by $({\cal Q}_{\bar
k})^{*\alpha \hat p}$ as well as by $({\cal Q}_{ k})^{\alpha  p} $. It has to
be
doubled since the CPT-conjugation adds the states where the role of Killing and
anti-Killing part of the spinors is reversed. This double-degeneracy of the
Clifford vacuum shows that black holes with opposite sign of central charges
behave as particle-antiparticle to each other. The spectrum of states generated
by
the action of $({\cal Q}_{\bar k})^{*\alpha \hat p}$  on the vacuum is
described as
follows. One set of states comes from
\begin{equation}
\Omega_{\bar k}\ , \quad  ({\cal Q}_{\bar k})^{*  \hat p}_\alpha \Omega_{\bar
k}\ , \quad \dots , \quad
({\cal Q}_{\bar k})^{*  \hat p_1}_\alpha   \dots  ({\cal Q}_{\bar k})^{*  \hat
p_{N/2}}_{\alpha }\Omega_{\bar k} \ .
\end{equation}
The Clifford vacuum state is a bosonic black hole with positive value of the
central charge, other states in this chain are black hole superpartners.
The set of CPT conjugate states is based on the analogous chain which starts
with the Clifford vacuum which is a black hole with the opposite sign of the
central charge.
The states are classified by the representations of $USp(N)\times SU(2)$ group.

The generators of both group  are constructed as bilinear combinations of
supercharges, where either the spinorial space-time indices  or the internal
ones are contracted.  In particular, the generator of the  $USp(N)$
transformations which labels the states of the given spin is
 \begin{equation}
s^{\hat m \hat n}  =-\,{i\over  2 \sqrt  |Z|}  \{ ({\cal Q}_{\bar k})^{  \hat
m}_
\alpha    , ({\cal Q}_{\bar k})^{  \hat n} _\beta \} \epsilon^ {\alpha \beta }
\ ,
\qquad \hat m,
\hat n = 1, \dots N\ .
\end{equation}
It generates the algebra of  $USp(N)$,
\begin{equation}
[s^{\hat m \hat n}, \; s^{\hat k \hat l}] = \Omega^{\hat m \hat k}  s^{\hat n
\hat l} + \dots
\end{equation}
Since we are going to use for gauge-fixing the split of the spinor defined by
the
supercharge of the background, one may expect that eventually the algebra
generating the full set of the BPS states will appear via Noether charges of
the
quantized string. We will show that  in the normal form (\ref{normal}) the
algebra
will be associated with the light-cone gauge and electrically charged black
holes.
The same algebra in  general  will be shown to represent the most   general
extreme black holes saturating the BPS bound.

\section {Duality-covariant
 gauges for the heterotic string}

We would like to consider the quantization of the heterotic string in the
four-dimensional  background of  extreme  black holes.
To be able to accommodate the black hole hair using the standard fields of the
Green-Schwarz string, which include the ten-dimensional Majorana-Weyl spinor
depending on the world-sheet coordinates $z,
\bar z$, we will use the form of  the constraints which Killing spinors
satisfy in this background, adapting them to the ten-dimensional form.
We may choose a commuting (anticommuting) $\kappa$-symmetry ghost of
some generation, which is a  ten-dimensional spinor, to  satisfy the constraint
\begin{equation}
\chi_{\bar k}\; C (z, \bar z) = {1 - \Gamma \over  2}   C (z, \bar z) =
0 \ .
\label{g.f.}\end{equation}
The numerical hermitean  matrix $\Gamma$ is defined by the properties of
the Killing spinors at asymptotic infinity of the target space. In our case it
is  defined  by the central charges of the background,
\begin{equation}
\Gamma = {Z \over |Z|} \ , \qquad Z^2 = |Z|^2 \ .
\end{equation}
We may use our constraint on the Killing spinor in the form which correspond to
that given  by Harvey and Liu
\cite{HL} and Sen \cite{Sen}  in their presentation of the form of the
Bogomolny
bound.
\begin{equation}
\Gamma = {i (\bar \lambda_0 - \lambda_0 ) \over 2 |Z|} \gamma^{ 0}
\sum_{a=4} ^{a= 9} \gamma^{ a} (Q_a + i\gamma_5 P_a) =
\Gamma^{\dagger} \ ,
\qquad
\Gamma^2 = 1 \ ,
\end{equation}
where
\begin{equation}
\lambda_0 = a_0 + i e^{-2\phi_0}
\end{equation}
is the value of the dilaton-axion complex scalar at infinity, far
away from the black hole. Six electric $Q_a$ and six magnetic $P_a$ charges
of the black hole satisfy the  conditions
\begin{equation}
\gamma^{[ab]} Q_a P_b =0 \
\label{ort}\end{equation}
and
\begin{equation}
 - e^{-4\phi_0} g^{00} g^{ab} (Q_a Q_b +
P_a P_b) = |Z|^2 \ .
\end{equation}
The 12 charges $Q_a, P_a$ can be  defined also
 via 28-dimensional charges $ (q_{\hat a })_ {el}  ,
\; (q_{\hat a })_{
mag}$ introduced by Sen \cite{Sen}
\begin{equation}
Q_a = E_a {}^{\hat a} ( q_{\hat a })_ {el} \ , \qquad  P_a = E_a {}^{\hat a}
(q_{\hat a })_{ mag} \ .
\end{equation}
The matrices $E_a {}^{\hat a}$ are defined by the non-vanishing expectations
values of the scalars of the four-dimensional theory (or, equivalently, by the
geometry of the compactified dimensions). Condition (\ref{ort}) was not spelled
explicitly in \cite{HL,Sen}. For us this condition is of great importance: in
this
form it reflects the critical orbit with $USp(N)$-symmetry, discussed in
Sect. 4. At the technical level, without (\ref{ort}) we
would
not
be able to get the required projectors in the presence of both electric and
magnetic
charges.

The Killing spinor  can be represented in the form
\begin{equation}
C (z, \bar z) = \chi_{ k}\;  C (z, \bar z) =  {1 + \Gamma  \over  2}   C (z,
\bar z) \ .
\end{equation}
Our new projectors  $\chi_{ \bar k}$, $\chi_{ k}$ indeed satisfy all the
requirements (\ref{proj}), since
\begin{eqnarray}
&&\left({ 1 + \Gamma  \over  2} \right) \;  \left({ 1 - \Gamma  \over
2}\right)
=0 \ , \\
&&\left({1 + \Gamma
\over  2}\right) ^2  = \left({1 + \Gamma  \over  2} \right) \ ,
\\
 && \left( {1 - \Gamma  \over 2}\right)^2  = \left({1 - \Gamma  \over  2}
\right) \ .
\end{eqnarray}

Under S- and T-duality transformations the central charges are
covariant. The corresponding covariant transformation of spinors makes our
constraint on $\kappa$-symmetry ghost duality covariant. Before discussing
the details of the covariant gauge-fixing, let us break both S- and
T-duality of the new class of gauges and reconstruct
the familiar class of gauges: the light-cone one and the generalized
light-cone gauge.

{\it Example 1 : light-cone gauge}

Our first example is a pure
electric $U(1)$ dilaton black hole. We choose the following hair
\begin{equation}
Q_9 = e^{2\phi_0} |Z|  \ , \qquad Q_4= \dots = Q_8 = P_4= \dots = P_9 = 0 \ .
\label{el1}\end{equation}
The black hole projector becomes a light-cone projector
\begin{equation}
\chi_{\bar k} = \left( {1 - \Gamma  \over 2}\right) =  \left( {1 - \gamma^0
\gamma^9
\over 2}\right) \ , \qquad \chi_{ k} = \left( {1 + \Gamma  \over 2}\right) =
\left( {1
+ \gamma^0
\gamma^9
\over 2}\right) \ .
\end{equation}

Thus in terms of the
ten-dimensional Majorana-Weyl spinor the electric solutions admit a Killing
spinor which satisfies the  light-cone constraint
\begin{equation}
(\gamma^- \gamma^+) \epsilon_k =  (\gamma^- \gamma^+) C (z, \bar z) =0 \ ,
\end{equation}
and the anti-Killing spinor satisfies equation
\begin{equation}
(\gamma^+ \gamma^-)  \epsilon_{\bar k} =0 \ .
\end{equation}
If we would choose a negative value of the electric charge
\begin{equation}
Q_9 = - e^{2\phi_0} |Z| \ ,  \qquad Q_4= \dots = Q_8 = P_4= \dots = P_9 = 0 \ ,
\qquad
\label{el2}\end{equation}
the constraint on the ghost would become the one for the anti-Killing spinor in
the previous choice,
\begin{equation}
(\gamma^+ \gamma^-)  C (z, \bar z)=0 \ .
\end{equation}

Thus in this example the
electric black hole hair breaks
 the ten-dimensional Lorentz symmetry $SO(1.9)$   down to
$SO(1.1)\times SO(8)$. We can consider the limit of our gauge-fixing function
(\ref{g.f.})  when the central charge vanishes, $|Z| \rightarrow 0$. This limit
exists and
has the same form $(\gamma^+ \gamma^-)  C (z, \bar z)=0$. Thus using the
light-cone gauge one can either consider the massive four-dimensional black
hole states or
massless four-dimensional states.

{\it Example 2 : generalized light-cone gauge}

Let us choose
\begin{equation}
Q_a = l_a e^{2\phi_0} |Z| \ ,   \qquad  P_4= \dots = P_9 = 0 \ .
\end{equation}
The constraint on the ghosts depends now on the
arbitrary six-dimensional vector $l_a$ satisfying the constraint $l^2 =1$:
\begin{equation}
(1-  \gamma^0 \gamma^a l_a )C (z, \bar z) =0 \ .
\end{equation}

This is a special choice of our generalized light-cone gauge when the vector
$n_{\mu} = (1,0,0,0, l_a)$, and $ n^2 =0$ due to the fact that $l^2 =1$.

{\it Example 3 :  magnetic gauge}

The quantization of superstring theory as well as of heterotic string theory
was
performed only in light-cone or generalized light-cone gauge. Therefore it was
widely believed  that the elementary excitations of string can be associated
only with electrically charged black holes. However we may change the gauge
now. Let us first consider the simplest  magnetic
$U(1)$ dilaton black hole.
\begin{equation}
P_4 = e^{2\phi_0} |Z| \ , \qquad Q_4= \dots = Q_9 = Q_5= \dots = P_9 = 0 \ .
\label{mag}\end{equation}

Now the  Killing spinor  and the $\kappa$-symmetry ghost
 in terms of the ten-dimensional Majorana-Weyl spinor  are constrained to
be chiral in the four dimensional Euclidean subspace, $\Gamma^5 = \gamma^1
\gamma^2 \gamma^3 \gamma^4 $,
\begin{equation}
(1 + \Gamma^5)\;\epsilon_k = (1 + \Gamma^5)\; C (z, \bar z)=0 \ ,
\end{equation}
and the anti-Killing spinor is anti-chiral,
 \begin{equation}
(1 - \Gamma^5)\epsilon_{\bar k} = 0 \ .
\end{equation}
Again, by changing the sign of the magnetic charge we have the
anti-chiral ghost.
Such split breaks the ten-dimensional Lorentz symmetry  to
$SO(1.5)\times SO(4)$.
This algebraic constraint has not been used before for the gauge-fixing of the
$\kappa$-symmetry.

{\it Example 4 : generalized magnetic gauge}

One can choose a more general magnetic charge with $SO(6)$ symmetry as
\begin{equation}
P_a = l_a e^{2\phi_0} |Z| \ .
\end{equation}
The ghost will satisfy the condition
\begin{equation}
 (1 + \gamma^1
\gamma^2 \gamma^3 \gamma^a l_a)\; C (z, \bar z)=0 \ .
\end{equation}

{\it Example 5 : electric-magnetic-axion-dilaton $U(1)$ black hole}

We choose
\begin{equation}
P_4^2 + Q_4^2  = e^{4\phi_0} |Z|^2  \ , \qquad
Q_5= \dots = Q_9 = P_5= \dots = P_9 = 0 \ .
\label{elmag}\end{equation}
The Killing spinor satisfies the following constraint
\begin{equation}
\left(1-{\gamma^0 \gamma^4  e^{-2\phi_0}(Q_4 + \gamma_5 P_4)\over |Z| }
\right) \; C (z, \bar z)=0 \ .
\end{equation}
This gauge was also never used before for the quantization of the heterotic
string.

\vskip 1 cm

In dealing with central charges of supersymmetry algebra related to
supersymmetric extreme black holes it is more convenient to use the chiral
basis  as described in Sec. 4. This basis is   associated with
the symplectic spinors which in a clear way shows how the black hole
multiplets form the representations of $USp(N)\times SU(2)$ algebra.
 In this basis the
truncation condition takes the form
\begin{equation}
{{\cal Z} \over |Z| } \; C(z, \bar z)=0  \ ,
\label{sympl}\end{equation}
where the numerical symplectic matrix ${\cal Z} $ is defined in eq. (\ref{Z}),
and the ghost forms a symplectic spinor. In a more detailed form the constraint
is
\begin{equation}
 \pmatrix{ {Z^{ij}\over  |Z|}  &   {\rm I}
\cr
\cr
-   {\rm I}  & { Z^{*ij}\over  |Z|} \cr } \pmatrix{ C^j _\alpha
\cr
\cr
C^{*\alpha  j}\cr } =  \pmatrix{ {Z^{ij}\over  |Z|} C^j_\alpha   + C^{*\alpha
i}
\cr
\cr
-C^i_\alpha  + { Z^{*ij}\over  |Z|} C^{*\alpha     i}\cr }=0 \ .
\label{detailed}\end{equation}
The advantage of using symplectic form of the constraint is that, e.g., the
second
line in the right-hand side of eq. (\ref{detailed}) can be obtained from the
first
one by multiplication on $Z^{ki}$. The black hole basis for supersymmetry which
was used in \cite{US} is very close to the one  which is used there. In
particular, the $SU(4)$ matrices $\alpha, \beta$ were used instead of  six
matrices $\gamma^a$. Thus, if we
know   the
antisymmetric matrix $Z^{ij}$, we can build the
symplectic matrix ${\cal Z}$ and have a symplectic spinor gauge fixing in the
form (\ref{sympl}). For example, pure electric solution with $P=0$ and positive
electric charge
$Z^{ij}= \alpha^3_{ij} Q/\sqrt 2$,  and with
\vskip -0.1 cm
\begin{equation}
Z^{ij}= \alpha^3_{ij} |Z| =  i \sigma_2\;    \pmatrix{|Z|&0&0&0 \cr 0
&|Z|&0&0\cr 0&0&|Z|&0\cr
0&0&0&|Z|\cr}  =  |Z|  \pmatrix{0&1&0&0 \cr -1&0&0&0\cr 0&0&0&1\cr
0&0&-1&0\cr}  \label{explalbe}
 \end{equation}
\vskip -0.1 cm
\noindent provides the following symplectic $8\times 8$ matrix for eq.
(\ref{sympl}):

\begin{equation}
\left( {{\cal Z} \over |Z| }\right)_{el}  =
 \pmatrix{ i \sigma_2 \;  {\rm I}
&   {\rm I}
\cr
\cr
-   {\rm I}  &  i \sigma_2 \;  {\rm I}}  \ ,
\label{Zel}\end{equation}

where  ${\rm I}$   is the unit $4\times 4$ matrix. If we would consider a pure
magnetic solution we would get  $Z^{ij}= \beta^3_{ij} P/\sqrt 2$,
\begin{equation}
Z^{ij}= \beta^3_{ij} |Z| =  i \sigma_2\;    \pmatrix{- |Z|&0&0&0 \cr 0 &-
|Z|&0&0\cr 0&0&|Z|&0\cr
0&0&0&|Z|\cr}  =  |Z|  \pmatrix{0&-1&0&0 \cr 1&0&0&0\cr 0&0&0&1\cr
0&0&-1&0\cr}  , \label{explalbe2}
 \end{equation}
and the corresponding symplectic $8\times 8$ matrix would be

\begin{equation}
\left( {{\cal Z} \over |Z| }\right)_{mag}  =
 \pmatrix{ i \sigma_2 \;   \pmatrix{-1&0&0&0 \cr 0 &-1&0&0\cr 0&0&1&0\cr
0&0&0&1\cr}
&   {\rm I}
\cr
\cr
-   {\rm I}  &  i \sigma_2 \;  \pmatrix{-1&0&0&0 \cr 0 &-1&0&0\cr 0&0&1&0\cr
0&0&0&1\cr}
}  \ .
\label{Zmag}\end{equation}
Thus indeed we see that the electric solution corresponds to the central charge
matrix in the normal form, with all eigenvalues equal, whereas the magnetic
solutions presents the central charge in the form related to the normal one by
some  $U(4)$ transformation.

We
have used in this section the ten-dimensional form of the constraints on
$\kappa$-symmetry to be able to display the relations between
known gauge-fixing conditions and the electric black holes (or
supersymmetric gravitational waves in ten-dimensional context). We have
also given examples of the magnetic and mixed electromagnetic gauges which
were never used for quantization before. We have also shown that the chiral
four-dimensional spinors, especially in symplectic form, are very convenient
way to
display dualities via the transformations of the central charges.


\section{ Central charges
in the flat space limit}

When the string is quantized in the BPS background, the central charges are
those of the background. To perform the quantization in the flat space
 we may consider different possibilities.  We may simply use the central charge
matrix of the background to gauge-fix the fermions as in eq. (\ref{sympl}) even
when the background is not there anymore. The second possibility is to consider
the limit of central charges going to zero,
\begin{equation}
\lim_{{\cal Z} \rightarrow  0} \left({{\cal Z} \over |Z| }\right)   C \equiv
\chi_{\bar k} C =0  \ ,
\label{sympllimit}\end{equation}
Such limit exists, we have shown examples in the previous section. In
particular when the background is pure electric or pure magnetic, the
constraint has the same for as the limit when the central charge goes to zero.
We may however take the following attitude. In the curved background with the
central charges
 the vacuum expectation value of the string variable (or zero mode,
$z,\bar z$-independent value)  $\Pi _z{}^\mu (z,
\bar z)$ is such as to reproduce the central charge in the supersymmetry
algebra.
The
ten-dimensional string variable
\begin{equation}
\Pi_{\bar z}{}^\mu (z, \bar z)= \partial_{\bar z} x^{\mu} - \bar \theta
\gamma^\mu \partial_{\bar z} \theta \ , \qquad \mu=0,\dots ,9\ ,
\end{equation}
consists of the four-dimensional part
$\Pi_{\bar z}{}^{\hat \mu} (z, \bar z) ,  \; \hat
\mu=0,\dots ,3$, and of the 6-dimensional part
$\Pi_{\bar z}{}^{a}(z, \bar z),  \; a=4,\dots
,9$. The extreme black hole background with specific values of the central
charge enforces  a non-vanishing value of the vacuum
expectation of the  string variables $\Pi$ of the following form:
\begin{equation}
\langle(\gamma_a {\cal C})_{\alpha \beta ij} \Pi_{\bar z}{}^{a} (z, \bar
z)\rangle \, =  \,
 \pmatrix{
Z^{ij}\epsilon_ {\alpha\beta}  & 0\cr
\cr
0   & Z^{*ij} \epsilon_{\dot \alpha \dot \beta} \cr } \ .
\end{equation}
Here   ${\cal C}$ is the charge conjugation matrix. The four-dimensional part
we take directly in the rest frame
be
\begin{equation}
\langle(\gamma_{\hat \mu}  {\cal C})_{\alpha \beta ij} \Pi_{\bar z}{}^{\hat \mu
} (z,
\bar z) \rangle =
\pmatrix{ 0  &  |Z| {\rm I} \cr
\cr
- |Z|  {\rm I}  &  0 \cr }  \ ,
\end{equation}
where ${\rm I}$ is the unit matrix.
Thus we define the string black hole  state  as the state with  vanishing
ten-dimensional
mass
of the state, however the four-dimensional mass is not vanishing
since the state has non-vanishing central charge.
  Indeed
we have
\begin{equation}
\langle(\gamma_\mu {\cal C} \;\Pi_{\bar z}{}^{\mu})\rangle \,
\langle(\gamma_\mu  {\cal C}  \; \Pi_{\bar z}{}^{\mu}) ^\dagger \rangle =
m_{10}^2 =0 \ ,
\end{equation}
however
\begin{equation}
\langle(\Pi_z{}^{a} \gamma_a {\cal C})\rangle \,
\langle(\Pi_z{}^{a} \gamma_a  {\cal C} ) ^\dagger \rangle =-  |Z|
^2 {\rm I} \ , \qquad \langle(\gamma_{\hat \mu} {\cal C} \;\Pi_z{}^{\hat
\mu})\rangle \,
\langle(\gamma_{\hat \mu}  {\cal C}  \; \Pi_z{}^{\hat \mu}) ^\dagger \rangle =
|Z|
^2 {\rm I} \ ,
\end{equation}
In this way we have reproduced the property of the centron BPS multiplet
discussed in \cite{K95} that the BPS state corresponds to a massless
ten-dimensional state but to a massive four-dimensional one.

To summarize, the string variable  $\Pi_z{}^\mu (z, \bar z) , \; \mu=0,\dots
,9$
 gets a non-vanishing $(z, \bar z)$-independent numerical value defined by the
central charge of the background.
\begin{equation}
\langle(\gamma_{ \mu}  {\cal C})_{\alpha \beta ij} \Pi_{\bar z}{}^{ \mu } (z,
\bar z)\rangle =
\pmatrix{ Z^{ij}  &  |Z| {\rm I} \cr
\cr
- |Z|  {\rm I}  &  Z^{*ij} \cr } \equiv {\cal Z} \ .
\label{spont}\end{equation}
Thus even if the BPS-background which supplies this matrix is absent we may
attribute the numerical values of the central charge matrix  to the
vacuum expectation of the string momenta. This again gives us a
constraint on
$\kappa$-symmetry ghost in the form  (\ref{sympl}).

The construction above suggest the following idea. One can rewrite the
classical GS
action for the heterotic string by using the string variables in the form
\begin{equation}
x^{\hat \mu}, \, \theta^i_\alpha, \, (\theta^i_\alpha)^*,  \, x^{ij}, \,
x^{*ij}, \qquad
{\hat \mu} = 0,1,2,3 \ , \quad i,j =1,2,3,4 \ , \quad \alpha=1,2 \ .
\end{equation}
The antisymmetric matrices $x^{ij} = ||x||$ which are the new bosonic
coordinates
of the compactified string, are defined as
\begin{equation}
(\gamma_a {\cal C})_{\alpha \beta ij} x^{a} (z, \bar z)  =   \pmatrix{
x^{ij}(z, \bar z)\epsilon_ {\alpha\beta}  & 0\cr
\cr
0   & x^{*ij}(z, \bar z) \epsilon_{\dot \alpha \dot \beta} \cr } \equiv X (z,
\bar z) \ ,
\end{equation}
and have the property
\begin{equation}
||x ||  || x^\dagger|| = (\sum_4^9 x^a x_a) \  {\rm I} \ .
\end{equation}

In terms of these variables the classical Green-Schwarz action for the
heterotic string has the global
$(z,
\bar z)$-independent
$U(4)$ symmetry under
which the fermionic string variables as well as bosonic variables $X$
transform.
The symmetry is best expressed in terms of the symplectic transformations in
the
form given in Sec. 3. The symplectic spinor is rotated as
\begin{equation}
\theta  \rightarrow T \theta \ , \qquad X\rightarrow TXT^T \ , \qquad T=
\pmatrix{
U   & 0\cr
\cr
0   & U^* \cr } \ .
\label{symplectic}\end{equation}
When the string action is considered in these variables, the generation of
central
charges matrix becomes a very natural step. This may lead to a reformulation of
the $N=4, N=8$ supersymmetry in a basis with central charge-type coordinates.
In particular, it was suggested in \cite{K81} to consider a $d=4$ superspace
with
additional coordinates for describing extended supergravities with hidden
supersymmetries. The full set of coordinates (for $N=8$ ) is
\begin{equation}
x^{\hat \mu}, \, \theta^I, \, \bar \theta^I, \,   t_{IJ}, \, \bar t^{IJ} \ ,
\qquad
{\hat \mu} = 0,1,2,3 \ , \quad  I,J =1,\dots ,8 \ .
\end{equation}
The bosonic coordinates $t_{IJ}, \, \bar t^{IJ}$ correspond to Cartan
antisymmetric tensors $x^{ij}, \, y_{ij}\ , i,j =1,\dots ,8$ which gives the
explicit
form of $E_7$ generators.  We  suggested to introduce the new vielbein
forms
$E_{ij}, \, \bar E ^{ij}$ in addition to the usual ones.  Those forms in curved
superspace at
$\theta=\bar \theta =0$ are defined by the scalar field matrix of Cremmer and
Julia \cite{CJ}
\begin{equation}
 \pmatrix{
U^{IJ}{} _{ij}   & \bar V^{IJij}  \cr
\cr
 V_{IJij} &  \bar U_{IJ}{}^{ij} \cr } \ .
\label{hidden}
\end{equation}
One can expect that the development of this direction will lead to the better
understanding of the role of central charges in supersymmetric theories. In
particular, the crucial property of all non-linear manifestly supersymmetric
on-shell invariants of $N=8$ supergravity is their $E_7$ symmetry and the fact
that they are build as the integrals over the full superspace
\cite{K81,HoweUl}.

The non-renormalization theorem for extreme black holes which was presented in
\cite{K92,US} had only one crucial requirement: fermionic isometries, which
make
the superfields covariantly independent on some fermionic coordinates. It seems
to become   clear now that if the  extreme black holes with manifest
$E_7$ symmetry\footnote{ Under manifest  $E_7$ symmetry we mean the
following. When the black hole hair (the values of the scalar field at infinity
and
the electric and magnetic charges of the black hole) undergoes the global
$E_7$ rotation, the total solution, as a function of three-dimensional space,
will rotate according to $E_7$. This property was demonstrated for manifestly
$SL(2,R)$ symmetric black holes in $N=4$ supergravity \cite{KO}. In all cases
the
classical symmetry groups are broken down in quantum theory to subgroups with
integer parameters only.}
will be
discovered as solutions of $N=8$ supergravity, they will represent the
U-duality symmetry of the spectra of quantized states of the superstring
theory. In addition, they will

i) form the most general background of the compactified to $d=4$ type II GS
string
theory in which the consistent truncation and quantization of
$\kappa$-symmetry is possible.

ii) these black holes will be subject to supersymmetric non-renormalization
theorem of the type described in  \cite{K92,US}.


\section{GS Superstring Path Integral in Black Hole Gauges}

For the time being, before the reinterpretation of string variables
responsible for accommodation of central charges is performed, we will study
the quantization of the GS heterotic string in the old variables ($ x^\mu,
\theta_\alpha ,\; \, \mu = 0,1, \dots ,9, \; \, \alpha = 1, \dots , 16$),  but
in
duality
covariant gauges.

The gauge fixed path integral in semi-light cone gauge $\gamma^+\theta=0
 $, $g_{\alpha \beta}= \rho  g_{\alpha \beta}^m$, where   $g_{\alpha
\beta}^m$ is some background metric, was presented in eq. (2.2) of \cite{KM} in
the
form\footnote{The semi-light cone quantization of the heterotic GS string was
performed by S. Carlip \cite{car} in a slightly different form.}
\begin{eqnarray}
&&\sum_{\rm topologies} e ^{-4\pi \chi \phi_0} \int_{moduli} \int D x^\mu
D\theta D\psi Db Dc (\mbox{Det} \;u_{\bar z})^{-4} \nonumber\\
&&\times \exp \left( - \int d^2 z (\partial_z
x^\mu
\partial_{\bar z} x^\mu + \bar \theta  \gamma^- \Pi_{\bar z}{}^+ \partial_z
\theta +L'_{\rm gauge}(\psi) + b\bar \partial c + \bar b \partial \bar c)
\right) \ ,
\end{eqnarray}
where $ u_{\bar z} = \partial_{\bar z} x^+ = \Pi _{\bar z}{}^+$.
In a generalized semi-light-cone gauge
$\chi_k (g. l.c.)= ({\slash \hskip -6.5pt {n}}\;  {\slash \hskip -9pt
{m}})$ the only difference would come in the
$\theta$ term in the action which will read
\begin{equation}
 \theta^T    U_{\bar z} \partial_z \theta
\label{spin}\end{equation}
 with
\begin{equation}
 U_{\bar z} (z,\bar z) = \chi_{k}{}^T {\cal C} {\slash \hskip -7pt {\Pi_{\bar
z}}}\,
\chi_{k}  \ ,
\end{equation}
and the local measure of integration is $(\mbox{Det}\, u_{\bar z})^{-4}$, which
 is equal to the inverse square root of
 the determinant of the maximum square invertible minor of the matrix $||U||$.

We
may rewrite this path integral now in the form where the Killing-anti-Killing
split of the spinor is realized in terms of the most general possible central
charge in $USp(4)$ critical orbit. Thus we consider the generic ``black hole
gauge''
\begin{equation}
\chi_{\bar k} \theta=0 \ , \qquad \theta = \theta_k \ ,
\label{bhgauge}\end{equation}
which means that only the Killing part of the anticommuting spinor $\theta$
propagate. In this case we have the same path integral as explained for the
generalized light-cone gauge, however with {\it any possible choice of the
projector }
$\chi_{\bar k}$ (see examples in Sec. 4) and not only the light-cone one. Thus
the
path integral is
\begin{eqnarray}
&&\sum_{\rm topologies} e ^{-4\pi \chi \phi_0} \int_{moduli} \int D x^\mu
D\theta D \psi Db Dc (\mbox{Det} \;u_{\bar z})^{-4} \nonumber\\
&&\times \exp \left( - \int d^2 z (\partial_z
x^\mu
\partial_{\bar z} x^\mu +  \theta^T   (\chi_{k}{}^T {\cal C} {\slash \hskip
-7pt
{\Pi_{\bar z}}}\,
\chi_{k})\partial_z
\theta +L'_{\rm gauge}(\psi) + L_{\rm rep. ghosts}) \right) \ .
\label{pathint}\end{eqnarray}
It was  explained in \cite{KM} that the local measure of integration provides
at least
formally the independence of the theory of the way in which local fermionic
gauge symmetry was fixed. This means that the part of the path integral given
by
\begin{eqnarray}
&&\int D x^\mu
D\theta D \psi Db Dc (\mbox{Det} \;u_{\bar z})^{-4} \nonumber\\
&&\times \exp \left( - \int d^2 z (\partial_z
x^\mu
\partial_{\bar z} x^\mu +  \theta^T   (\chi_{k}{}^T {\cal C} {\slash \hskip
-7pt
{\Pi_{\bar z}}}\,
\chi_{k})\partial_z
\theta +L'_{\rm gauge}(\psi) + L_{\rm rep. ghosts}) \right) \
\label{invar}\end{eqnarray}
is invariant under the change of the gauge conditions.
This gauge is  a unitary gauge for the fermionic
symmetry (all fermionic ghosts are not propagating). The local measure of
integration is exactly the contribution of the second class constraints as
predicted in \cite{F,BV}. The class of gauges which were considered before
and the transformations from one to another  did not involve any changes
of the vacuum expectation value of the dilaton which controls the loop
expansion. Therefore for this class of gauges the fact the that integral in
(\ref{invar}) is invariant by construction is sufficient to claim that the
total
path integral including the loop integrations
\begin{equation}
\sum_{\rm topologies} e ^{-4\pi \chi \phi_0} \int_{moduli}
\label{loops}\end{equation}
is gauge invariant. Now we are considering  more general class of gauges
 which are related by S-duality transformations from
one gauge to another. The partition function on the torus where $e ^{-4\pi \chi
\phi_0} =1$ is now duality invariant by construction. However if we are
interested in partition functions for different topologies, we have to take
into account that S-duality will act on  the string coupling constant as
follows;\begin{equation}
(e ^{-2 \phi_0})' = \left (c (a_0 + i e ^{-2 \phi_0}) + d\right) ^{-1}
\left (c (a_0 - i e ^{-2 \phi_0}) + d\right) ^{-1} e ^{-2 \phi_0} \ ,
\label{diltransf}\end{equation}
where $a_0$ is the value of the axion field at infinity and $c,d$ are some
integers. The part of the path integral given in eq.
(\ref{invar}) is invariant. However each term in the path integral in
(\ref{pathint}) with the non-vanishing Euler number $\chi$ transforms
when we use the
full
$SL(2,Z)$ transformation to change a constraint on the spinor\footnote{We
will consider in detail the S-duality covariant gauges in Sec. 8.}and on the
string coupling constant.
 Thus when duality transformation includes the dilaton whose vacuum
expectation value plays the role of the string coupling constant, each term in
the
 Green-Schwarz path integrals given in eq. (\ref{pathint}) is covariant rather
than
invariant: the change of the gauge has to be followed by the corresponding
change of the coupling constant. There is a puzzling resemblance here
with the observation about S-duality due to Witten \cite{W95}.  He found that
the partition function on a general four-manifold is not modular invariant but
transforms as a modular form of the weight depending on the topology of the
manifold. Using the fact that our partition function consists of the invariant
part given in eq. (\ref{invar}) and using the $SL(2,Z)$ transformation
(\ref{diltransf}) one can see that for each topology our partition function
transforms as a modular form of the weight depending on the topology of the
manifold.

The expansion in topologies  makes the understanding of S-duality more
complicated and perhaps the better way to proceed is to
use the first quantization, suggested above only to get the elementary string
states, which are duality invariant, according to eq. (\ref{invar}). The next
step would be  to construct the BRST
operator for the first quantized theory. The resulting second quantized theory
may have better
way of realizing S-duality and may give us a possibility to avoid the loop
expansion in the form
(\ref{loops}).

Therefore for the moment we will
concentrate on the part of the path integral given in eq. (\ref{invar}) which
has
a clear behavior  under the change of the gauge conditions including the
S-duality type.
For example in the pure magnetic gauge where $\chi_k = {1\over \sqrt 2}(1-
\Gamma^5)$ the fermionic part of the action is
\begin{equation}
 {1\over 2} \bar \theta   ( 1- \Gamma^5) {\slash \hskip -7pt {\Pi_{\bar z}}}\,
( 1-
\Gamma^5)
\partial_z \theta = {1\over 2} \bar \theta   ( 1- \Gamma^5) \left( \gamma_0
\Pi_{\bar z}{}^0 + \sum_{a=5}^{a=9}  \gamma_a \Pi_{\bar z}{}^a \right)
\partial_z \theta \ .
\end{equation}
Thus the kinetic term of the fermionic variables depends on $SO(1,5)$ vector
$\Pi^0, \Pi^a$, whereas the spinor is chiral in $SO(4)$. The local measure for
the magnetic solution is
\begin{equation}
\left[\mbox{Det} \Bigl ( -(\Pi_{\bar z}{}^0)^2 + \sum_{a=5}^{a=9} (\Pi_{\bar
z}{}^a)^2
\Bigr)\right]^{-2} \ .
\end{equation}

 There are different ways to proceed with this action. In \cite{KM} we
  performed some change of variables after which we got the
$SU(4)$-form of the path integral. The analogous change of variables can be
performed starting with any gauge of the type (\ref{bhgauge}). This results is
\begin{eqnarray}
&& \int D x^\mu
D \eta_{\bar z}{}^i D\theta_i  Db Dc  \nonumber\\
&&\times \exp \left( - \int d^2 z (\partial_z
x^\mu
\partial_{\bar z} x^\mu + \sum_{i=1}^{i=4}\eta_{\bar z}{}^i
\partial_z
\theta_i  +L'_{\rm gauge}(\psi) + L_{\rm rep. ghosts}) \right) \ .
\end{eqnarray}
This form of the GS superstring path integral shows in a clear way that
the idea that the elementary string excitations have to be associated only with
electric black holes is based on the  light-cone gauge quantization. In the
generic class of gauges the elementary string excitations cannot be qualified
as
electric black holes: they are given by generic black holes. We have presented
the string partition function in the form in which there is no dependence left
on the choice of the constraint on the spinor.Therefore there is no difference
whether we have started with electric-type constraint $\gamma^+ \theta=0$ or
magnetic-type constraint $(1+ \Gamma^5)\theta =0$.
One can claim on the basis of this construction that  the elementary string
excitations
is invariant under duality transformations as the
soliton configurations solving the
classical equations of motions and as the Bogomolny bound.

Another way to proceed with the path integral in the semi light-cone gauge was
suggested in \cite{PN}. These authors  rescaled  the 8-dimensional Killing
spinors
$\theta$ without breaking it into two $SU(4)$ spinors. These procedure seems to
be more suitable for dealing with anomalies. At this stage we would prefer to
postpone the issue of anomalies and just work out the black-hole class of
gauges
which contains the light-cone gauge as a subclass. We hope that the situation
with conformal and gauge fermionic symmetry anomalies will be studied later
along the lines of \cite{car,KM,PN}.

One more comment about the black hole gauges is in order. If we would choose
the light-cone gauge also for the bosonic variables of the string, i.e.
\begin{equation}
x^+ = \tau P^+\ , \qquad \partial_{\bar z} x^+ = P^+ \ ,
\end{equation}
we would have to identify the variable $P^+$ with the mass of the black hole.
Indeed, in our picture the origin of the light-cone gauge is traced back to the
central charges. However, they appear via the zero modes of the string momenta.
Therefore
\begin{equation}
\langle \Pi_{\bar z}{}^0 +\Pi_{\bar z}{}^9\rangle =P^+ = 2 |Z| = 2 M \ ,
\end{equation}
 where $M$ is the mass of the
black hole.

Taking into account that  $P^+$ plays such an important
role in the Green-Schwarz string field theory based on the light-cone first
quantization one may hope that our picture may lead to string field theory
describing the duality symmetric interactions of extreme black hole multiplets.

{\it Manifestly supersymmetric black-hole gauge}

Manifestly supersymmetric gauge was suggested in \cite{K87} with  the purpose
to keep linear realization of supersymmetry. The algebraic constraint was
imposed on the first generation of fermionic ghosts, whereas the gauge for the
$\theta$-variable was chosen to contain a derivative. We may present now
the gauge-fixed action for  manifestly supersymmetric  black hole gauges.
The constraint on the first generation of ghosts is defined by the central
charge
matrix. We will use it in the form $\chi_{\bar k} C(z, \bar z)=0$. The
gauge-fixed action according to eqs. (38) from \cite{K87} and eq. (3.4) of
\cite{KM} is
\begin{equation}
{\cal L}_{cl} + \bar \pi_{\bar z}  \chi_{\bar k} \partial_z \theta +
(\partial_z\bar
C_{\bar z})
\chi_{\bar k}  {\slash \hskip -7pt {\Pi_{\bar z}}}\,
\chi_{k} C_{z} \ ,
\end{equation}
where the propagating  FP ghosts $\bar
C_{\bar z}, \, C_{z}$ are commuting whereas the propagating NK ghosts $\bar
\pi_{\bar
z}$ are  anticommuting. By performing some change of variables and by adjusting
the local measure of integration one can prove that the physical states are
independent of the choice of the fermionic gauge-fixing.


\section{S-duality symmetric family of black hole gauges}

The basic
new feature of the quantization which we propose is the use of the algebraic
constraint on Killing spinor of the BPS background. The simple and
universal form of the constraint is given in eq. (\ref{sympl}). When
the background  undergoes any duality transformation, the Killing
spinor and the algebraic constraint on Killing spinor transform in a way which
reflects  {\it the
symmetry under dualities of   equations of motion including   fermions}.

In this section we
would like to study the new quantization for the special case  of
the  axion-dilaton black holes in manifestly S-duality symmetric form
\cite{KO}. For this configuration we know exactly how background
(in our example the superspace, whose bosonic part consists of
axion-dilaton black holes) transforms under S-duality and what happens with
the constraint on Killing spinor. After the detailed analysis of this
configuration
we will reformulate in the next section the information available about the
S-duality covariant gauges to the form suitable for the generalization to the
U-duality symmetric gauges.

The axion-dilaton family of black holes \cite{KO} has a feature which justifies
the word family.  We consider  the solution  in a form in which it is
characterized by some generic  values  of the the axion-dilaton field
and
electric and magnetic charges.  After the S-duality transformation the solution
keeps the same functional form. It is important that one considers the generic
values of the black hole hair and not the exceptional cases like pure magnetic
or
pure electric solutions, for example. In such special cases, as it was
demonstrated in
\cite{TW,TO93}, one starts with pure electric solutions and after S-duality
transformation one gets a new solution which is characterized by the electric
as
well as magnetic charge and by some value of the axion field, which was not
present in the original pure electric dilaton black hole. However, after the
manifest S-duality symmetric form of the solutions is found one does not
generate
new solutions by performing an additional S-duality transformation, they are
all
there, in the family. For simplicity we will consider a $U(1)$ axion-dilaton
black hole \cite{KO} with only one vector field, which has both magnetic and
electric charge\footnote{In this section we are using the notation of \cite{KO}
unless otherwise specified.}.

 The solution has the following form:
\begin{eqnarray}
ds^{2}           & = & e^{2{\cal U}}dt^{2}-e^{-2{\cal U}}d\vec{x}^{2}\;
, \hspace{1cm}
                                e^{-2{\cal U}}(\vec{x})= i
({\cal H}_{2}(\vec{x})\;
                                 \overline{{\cal
                                H}}_{1}(\vec{x}))  -  {\cal H}_{1}(\vec{x})\;
                                 \overline{{\cal
                                H}}_{2}(\vec{x})\; ,
                                    \nonumber \\
                       \nonumber \\
A^{t} (\vec{x})       & = &(Q+iP) {\cal
                                H}_{2}(\vec{x})+c.c. \; ,\hspace{1cm}
\lambda(\vec{x})  =  a( \vec{x}) + i  e^{-2 \phi  (\vec{x}) }  =\frac{{\cal
H}_{1}(\vec{x})}{{\cal
                         H}_{2}(\vec{x})}\; ,
                        \nonumber \\
\nonumber \\
\tilde{A}^{t}(\vec{x}) & = & -(Q+iP) {\cal
                                 H}_{1}(\vec{x})- c.c. \; ,
\end{eqnarray}
where ${\cal H}_{1}(\vec{x}), {\cal H}_{2}(\vec{x})$
are two complex harmonic functions (for simplicity we are considering a
one-black-hole solution)
\begin{eqnarray}
{\cal H}_{1}(\vec{x}) & = & \frac{e^{\phi_{0}}}{\sqrt{2}}
                            \{\lambda_{0} +
\frac{\lambda_{0}M+
                            \overline{\lambda}_{0}
                            \Upsilon}{|\vec{x}
                            |}\} \equiv\, v_1 + {{\cal M}_1 \over |\vec{x}|}\;
,
                                      \nonumber \\
\nonumber \\
{\cal H}_{2}(\vec{x}) & = & \frac{e^{\phi_{0}}}{\sqrt{2}}
                            \{1+\frac{M+
                            \Upsilon}{|\vec{x}
                            |}\} \hskip 1 cm \equiv\, v_2 + {{\cal M}_2 \over
|\vec{x}|}  \; .
\label{harmonic}\end{eqnarray}
where $\lambda_{0} \equiv \lim_{|\vec x| \rightarrow \infty}
\lambda(\vec{x})$.

 The S-duality transformation on the half supersymmetric bosonic background is
given by the fractional transformation on the complex scalar $\lambda(x)$:
\begin{equation}
\lambda' (x) = {a\lambda(x) + b \over c \lambda(x) + d} \ .
\end{equation}
Here the $SL(2, Z)$ matrix  $\Lambda$ is

\begin{equation}
\Lambda = \left |\matrix{
a & b\cr
c & d \cr
}\right |  \ , \qquad \Lambda^{-1} = \left |\matrix{
d & -b\cr
-c & a \cr
}\right | \ , \qquad \det \Lambda =1 \ ,
\end{equation}
where $a,b,c,d$ are some real integers.

We will present a useful form of this solution where only the $SL(2, Z)$
doublets enter. This will be helpful for clarifying the transformation property
of the gauge-fixing condition under  the S-duality. This form will be also
suggestive for the U-duality.
Consider the following harmonic matrix:

\begin{equation}
\partial_i  \partial_i   V(x) =0 \ ,\qquad V(x) = \pmatrix{
{\cal H}_{2}(\vec{x})  & - {\cal
H}_{1}(\vec{x}) \cr
\cr
\overline  {\cal H}_{2}(\vec{x}) & - \overline {\cal H}_{1}(\vec{x}) \cr
} \ , \qquad \det V(x)  = i e^{-2 {\cal U}(x) } \ .
\end{equation}
Under S-duality this matrix transforms as

\begin{equation}
V' (x) = h \,V(x) \Lambda ^{-1} \ ,
\end{equation}
where

\begin{equation}
h = \pmatrix{
U & 0 \cr
\cr
0 & U^* \cr
} , \qquad U= {|S_0| \over S_0} \ , \qquad S_0 \equiv c\lambda_0 +d \ .
\end{equation}
This transformation is known to be a   compensating  $U(1)$ transformation
which supports the choice of the
 local $U(1)$  gauge fixing under the
$SL(2,Z)$ transformation.

We can also use two harmonic doublets

\begin{equation}
\left (\matrix{
{\cal H}_{2} \; ,  -{\cal H}_{1} \cr
}\right )' = {|S_0| \over S_0} \left (\matrix{
{\cal H}_{2}  \; , -{\cal H}_{1} \cr
}\right ) \Lambda^{-1} \ ,
 \qquad
\left (\matrix{
{\cal H}_{1}\cr
{\cal H}_{2}\cr
}\right )' = {|S_0| \over S_0} \, \Lambda \left (\matrix{
{\cal H}_{1}\cr
{\cal H}_{2}\cr
}\right ) \ .
\end{equation}

The vector fields are also organized in doublets. We have  a harmonic doublet
potential

\begin{equation}
\partial_i  \partial_i  {\cal A} =0 \ ,  \qquad {\cal A} = \left (\matrix{
A^t (x)\, ,   \tilde A^t(x) \cr
}\right ) \  .
\end{equation}
Under  S-duality it transforms as

\begin{equation}
{\cal A} ' = {\cal A}  \, \Lambda^{-1} \ .
\end{equation}

The vector field strength also can be presented in the doublet form:

\begin{equation}
 {\cal F} \equiv \left (\matrix{
F_{tr}  \; ,  -i   \tilde F_{tr} \cr
}\right )  = {1\over |\vec x |^2 } \left (\matrix{
q \; ,   \tilde p \cr
}\right ) \ , \qquad {}^* {\cal F}  \equiv \left (\matrix{
{}^* \tilde F_{tr} \cr
-i {}^* F_{tr} \cr
}\right ) = {1\over |\vec x |^2 }  \left (\matrix{
\tilde q \cr
p\cr
}\right ) \ .
\end{equation}

Those two doublets transform as follows under S-duality:

\begin{equation}
 {\cal F}' =  {\cal F}  \, \Lambda^{-1} \ ,
 \qquad {}^* {\cal F} ' =   \Lambda\,  {}^* {\cal F} \ .
\end{equation}

Our matrix $V$ consists of the value of this matrix at infinity when $ |\vec x
| \rightarrow \infty$ and of the ${1\over |\vec x | } $ part of this matrix:

\begin{equation}
 V(x) = v + {1\over |\vec x | }  {\cal M} \ ,\hskip 1 cm  v   \equiv  \pmatrix{
v_{2}  & - v_{1} \cr
\cr
\bar v_{2} & - \bar v_{1} \cr
}  ,  \hskip 1 cm  {\cal M}  \equiv  \pmatrix{
{\cal M}_{2}  & - {\cal
M}_{1} \cr
\cr
\overline {\cal M}_{2} & - \overline {\cal M}_{1} \cr
}   .
\end{equation}
The matrices  $v$ and ${\cal M} $ transform as follows:
\begin{equation}
v '  = h \,v  \Lambda ^{-1} \ , \hskip 2 cm {\cal M}  = h \, {\cal M}  \Lambda
^{-1} \ .
\end{equation}
Consider the following product of doublets:
\begin{equation}
(v \; {}^* {\cal F})' = h  (v \; {}^* {\cal F}) \ .
\label{prod}\end{equation}
Such products transform only in terms of the compensating $h$ transformation.
They will be useful for the form of the gauge-fixing which will transform
under $SL(2,Z)$ only in terms of the $h$ matrix.  Now we can build the
combination suitable for exhibiting S-duality covariant form of the
superstring
$\kappa$-symmetry gauge-fixing.

The Killing spinor admitted by the   axion-dilaton black holes was found
by  Ort\'{\i}n in  \cite{TO95}:
\begin{equation}
\epsilon _{I} (\vec{x}) = e^{{1\over 2} {\cal U}(\vec{x})}  \left(\frac{ {\cal
H}_{2}(\vec{x})}{\overline{\cal
                         H}_{2}(\vec{x})} \right)^{1\over 4}\epsilon _{I(0)}  \
{},
\end{equation}
where  $\epsilon _{I(0)}$ is the value of the chiral part of the Killing spinor
at
infinity, satisfying an algebraic constraint which halves the spinor.
We will study this constraint both in
the doublet  form
as well as in the form in which only the $h$ part of the symmetry is relevant.

The doublet form of the constraint is
\begin{equation}
 \chi_{\bar k} \epsilon = 0  \hskip 1 cm \Longrightarrow    \hskip 1 cm  \left
(\matrix{
i{\cal M}_1 \cr
i{\cal M}_2\cr
}\right ) \epsilon _{I(0)} - \left (\matrix{
\tilde q \cr
p\cr
}\right )_{IJ} \gamma^0 \epsilon^J_{(0)} =0 \ .
\label{tomas}\end{equation}

Under S-duality transformations the chiral part of the Killing  spinor
transforms as follows:
\begin{equation}
\left (\epsilon _{I} (\vec{x})\right )'  = e^{{1\over 2} {\cal
U}(\vec{x})}\left[ \left ( \frac{{\cal H}_{2}(\vec{x})}{\overline {\cal
                         H}_{2}(\vec{x})} \right )^ {1\over 4} \right]'  \left
(\epsilon _{I(0)}\right )'  = e^{{i\over 2} \arg S} \epsilon _{I} (\vec{x}) \ ,
\end{equation}
where
\begin{equation}
 S (\vec{x}) \equiv c\lambda (\vec{x}) +d = c \, \frac{{\cal
H}_{1}(\vec{x})}{{\cal
                         H}_{2}(\vec{x})} + d \ .
\end{equation}

Under S-duality the constant part  of Killing spinors  transforms in terms of
the asymptotic value   of    $S(\vec{x})$   at  ${|\vec{x}| \rightarrow
\infty}$ which is equal to $ S_0 =c\lambda_0 +d $:
\begin{equation}
(\epsilon _{I(0)})  ' = e^{{i\over 2} \arg S_0} \epsilon _{I(0)} \ , \qquad
(\epsilon _{(0)}{}^J)  ' = e^{- { i\over 2} \arg S_0} \epsilon _{(0)}{}^J \ .
\end{equation}
Taking into account the transformation of doublets above we find that the
Killing spinor constraint in the doublet form transforms as follows:
\vskip -.4cm
\begin{equation}
 \left (\left (\matrix{
i{\cal M}_1 \cr
i{\cal M}_2\cr
}\right ) \epsilon _{I(0)} - \left (\matrix{
\tilde q \cr
p\cr
}\right )_{IJ} \gamma^0 \epsilon ^J_{(0)}\right  )' \nonumber\\
\nonumber\\
 =   e^{- { i\over 2} \arg S_0}   \; \left |\matrix{
a & b\cr
c & d \cr
}\right |    \left (\left (\matrix{
i{\cal M}_1 \cr
i{\cal M}_2\cr
}\right ) \epsilon _{I(0)} - \left (\matrix{
\tilde q \cr
p\cr
}\right )_{IJ} \gamma^0 \epsilon ^J_{(0)} \right ) =0 \ .
\label{Stransform}\end{equation}
\vskip -.2cm
An alternative   form of the Killing constraint can be obtained by multiplying
eq.
(\ref{tomas}) by the doublet $( \bar v_2 \; , - \bar v_1)$ as suggested by eq.
 (\ref{prod}).
Indeed we may simplify things by using the fact that
\vskip -.4cm
\begin{equation}
( \bar v_2 \; , - \bar v_1) \times \left (\matrix{
i{\cal M}_1 \cr
i{\cal M}_2\cr
}\right ) = -M \ ,
\hskip 2 cm
( \bar v_2 \; , - \bar v_1) \times  \left (\matrix{
\tilde q \cr
p\cr
}\right ) = {1\over \sqrt 2} (Q+iP) \ .
\end{equation}
\vskip -.2cm
The Killing spinor constraint (\ref{tomas}) after multiplication by  $( \bar
v_2 \; , - \bar v_1)$ becomes
\begin{equation}
 M \epsilon _{I(0)} + {1\over \sqrt 2} (Q+iP)
_{IJ} \gamma^0 \epsilon ^J_{(0)} =0 \ .
\end{equation}
In this form it transforms as
 \begin{equation}\left (
M \epsilon _{I(0)} + {1\over \sqrt 2} (Q+iP)
_{IJ} \gamma^0 \epsilon ^J_{(0)}\right)'  =  e^{ i \arg S_0} \left(
M \epsilon _{I(0)} + {1\over \sqrt 2} (Q+iP)
_{IJ} \gamma^0 \epsilon ^J_{(0)}\right) =0 \ ,
\label{Scontracted}\end{equation}
and we have taken into account that $(Q+iP)' = e^{  {i\over 2}  \arg S_0}
(Q+iP)$.

We  considered a solution with only one vector field, i.e. our choice of
$(Q+iP)
_{IJ} $ was $\alpha^3_{IJ} ((Q+iP)
$ where   $\alpha^3_{IJ}$ is one of the $SU(4)$ matrices $\alpha, \beta$. In
more general situation we would have $(Q+iP)
_{IJ} = \alpha^n_{IJ} (Q+iP)_n + \beta^{\tilde n} ((P+iQ)_{\tilde n}$ where $
n, {\tilde n} = 1,2,3.$ In six-dimensionally covariant form\footnote{There is a
factor of $\sqrt 2$ difference in the definition of charges used in \cite{KO}
and \cite{HL}.} we would have the
constraint defined in notation of Sec. 4:
\begin{equation}
  \chi_{\bar k} \epsilon = 0  \hskip 2 cm \Longrightarrow  \hskip 2 cm \epsilon
- \gamma^0 \gamma^a (Q_a + i\gamma^5 P_a) \epsilon  =0 , \qquad a=4, \dots , 9.
\end{equation}
Under S-duality the constraint in this form transforms as
 \begin{equation}
  \left(  \epsilon - \gamma^0 \gamma^a (Q_a + i\gamma^5 P_a) \epsilon \right)'
=
e^{- { i  \gamma^5 \over 2} \arg S_0}  \left(  \epsilon - \gamma^0 \gamma^a
(Q_a + i\gamma^5 P_a) \epsilon \right) =0 \ .
\end{equation}
A nice form of this transformation comes out if we use symplectic notation
for the constraint on symplectic Killing spinor $C$ as given in the form

\begin{equation}
 \left({ {\cal Z} \over |Z| } C  \right)' =  h_{sp}  \left( {{\cal Z} \over |Z|
} C \right) =
0 \ , \qquad |Z|' = |Z| \ ,
\label{Ssympl}\end{equation}
where

\begin{equation}
h_{sp}  = \pmatrix{
U^{1/2}  & 0 \cr
\cr
0 & (U^*)^{1/2}  \cr
} ,  \qquad U^{1/2} = (S_0)^{1\over 4}  (\bar S_0 )^{-{1\over 4}} = e^{ {
i\over 2} \arg S_0} \ , \qquad S_0 \equiv
c\lambda_0 +d \ .
\label{mod}\end{equation}

It is important to stress that that if in the flat background we have chosen a
specific
gauge-fixing condition (truncation) and afterwards have decided to put the
quantized string in the background, we could do it only under condition that
the
algebraic constraint on the ghost defines the algebraic constraint on the
Killing spinor of the half-supersymmetric background. For example the string
quantized in the light-cone gauge can be placed in the electric black hole
background, but the string quantized in the magnetic gauge can be placed only
in the magnetic background. The string quantized in the electro-magnetic gauge
can be placed in the electro-magnetic black hole background. That is why we
call those gauges black hole gauges.

An additional useful way  to summarize this section is to use Witten's
definition of the modular forms of the weight  $(u,v)$  \cite{W95} if the
expression transforms as
\begin{equation}
F' =  (c\lambda_0 +d)^u \; (c\bar \lambda_0 +d)^v F \ .
\end{equation}
Our new black hole gauge conditions are modular forms of the weight $ \left(
\pm{1\over 4}, \, \mp {1\over 4}\right)$ for the left-handed (upper sign) or
right-handed (lower sign)
parts of spinorial gauge conditions, see eq.  (\ref{mod}).


\section{U-duality covariant class of  gauges}

Hidden symmetries of supergravities, restricted to the subgroups with integer
parameters  to provide the black hole
charge quantization, are realized in extreme black holes solutions.
In this respect {\it extreme black holes are as good for realizing dualities as
are quarks for realizing the fundamental representation of} $SU(3)$. In type II
strings or in the supermembrane case we may look for a general class of
duality-symmetric gauges although we do not know yet a black hole solution
with one half of unbroken supersymmetry and $E_7$ (or $SL(8,R)$)-covariant
black hole hair. However, one can use the information available about the
hidden symmetry of
$N=8$ supergravity
\cite{CJ}. We may just use the fact that it is known how the hidden symmetries
act on fields, in particular, on spinors. In symmetric gauge $N=8$
supergravity has an $SO(8)$ symmetric form. The scalar matrix  $V$ in this
form satisfies the condition
\begin{equation}
V =  V^\dagger \ .
\end{equation}
This condition is provided by the $SU(8)$ gauge transformation. The theory can
be formulated in the inhomogeneous coordinates of ${E_7 \over SU(8)}$. In these
coordinates  $Y_{AB,CD},  \, A,  \dots , D = 1, \dots ,8$ are the scalar fields
of the theory. If we would know the most general black hole solutions they
would probably give us those scalar fields as some functions of the space
coordinates $\vec x$, and the metric, the vector fields etc. would be adjusted
to the scalar matrix expectation value (values at infinity) as well as scalar
charges,  which will form the ${1\over |\vec x|}$ part of this matrix.
Thus as  different from the previously studied case of S-duality we do not
have the
solution available which exhibits this hidden symmetry, i.e. we do not know yet
the  configuration which solves classical equations of motion of $N=8$
supergravity in a form in which the solution has manifest U-duality. However
the general structure of the theory suggests that the algebraic constraint on
the Killing spinor
will undergo a transformation generalizing that of S-duality, which is
described above.

Let us first reformulate the information about the S-duality described above in
the form suitable for the generalization to U-duality.

The action of $U(1,1)$ (complex version of  $SL(2,R)$) on spinors of $N=4$
supergravity interacting with matter with the local $U(4)$ gauge symmetry fixed
was known since 1977\, \cite{CSF}. The scalars $Y(x)$ are related to the
axion-dilaton field $\lambda(x)$ described above.  For some arbitrary parameter
$\zeta$ (which is equal  to 1 in the absence of matter) the scalar matrix  may
be
chosen in the symmetric gauge
$V=V^\dagger$:
\vskip -.3 cm
\begin{equation}
V\left (Y(x)\right )  =
\pmatrix{
 {1\over \sqrt { 1- Y \bar Y}} &  {Y\over \sqrt { 1- Y \bar Y}} \cr
\cr
 {\bar Y \over \sqrt { 1- Y \bar Y}} &  {1\over \sqrt { 1- Y \bar Y}} \cr
}\ ,
\end{equation}
were
$Y  =  \zeta^{1\over 2} \left( {1+i\lambda (x) \over 2}\right) $.

When the scalars are subject to fractional transformation of $U(1,1)$,
\begin{equation}
Y = {aY' +\bar c \over cY' + a} \ , \qquad |a|^2 - |c|^2 = 1 \ ,
\end{equation}
the gravitino  has to transform as follows:

\begin{equation}
\Psi_\mu = \exp ^{{i \zeta\over 4 } \gamma_5 \theta(Y', \bar Y' )}
 (\Psi_\mu)' \ ,
\end{equation}
where
\begin{equation}
\exp ^{i \theta(Y', \bar Y') } =  {cY' +\bar a \over \bar c \bar Y' + a} \ .
\end{equation}
The Killing spinor, being defined as a zero mode of the equations
\begin{equation}
(\delta_{susy} \Psi_\mu)  = \exp ^{{i \zeta\over 4 } \gamma_5
\theta(Y', \bar Y' )} (\delta_{susy} \Psi_\mu)' =0 \ ,
\end{equation}
obviously has to transform under duality transformation $SU(1,1)$ as follows:
\begin{equation}
(\hat \nabla _\mu)\epsilon_{k}   = \exp ^{{i \zeta\over 4 } \gamma_5
\theta(Y', \bar Y' )} \left((\hat \nabla _\mu)\epsilon_{k}\right)' =0 \ .
\end{equation}
 This is all we need to show that our choice of the gauge-fixing  defined
by the Killing spinors of the background is duality covariant. Available black
hole solutions gave us examples of such rotations of the Killing spinors. In
particular, the Killing spinor admitted by the  axion-dilaton black holes in
symplectic form is
\vskip - .3 cm
\begin{equation}
\left (\matrix{
\epsilon (\vec{x})\cr
\cr
\epsilon^*(\vec{x}) \cr
}\right )= e^{{1\over 2} {\cal U}(\vec{x})}  \pmatrix{
  \left(\frac{ {\cal H}_{2}(\vec{x})}{\overline {\cal
                         H}_{2}(\vec{x})}\right) ^{1\over 4} & 0 \cr
0 &  \left (\frac{\overline  {\cal H}_{2}(\vec{x})}{{\cal
                         H}_{2}(\vec{x})}\right)^{1\over 4}  \cr
}
\left (\matrix{
\epsilon\cr
\cr
\epsilon^* \cr
}\right )_{(0)}  ,
\end{equation}
\vskip - .3 cm
\noindent where $\epsilon_{0}$ is the constant $x$-independent part of the
Killing spinor
(its value at  $|\vec x| \rightarrow \infty$).

We would like to stress that the compensating $U(1)$ transformation
acting on the Killing spinors of the background is neither global
nor local: it is {\it rigid}. Indeed,  the $h_{sp}$ matrix in our example is
\vskip - .3 cm
\begin{equation}
\left (\matrix{
\epsilon (\vec{x})\cr
\cr
\epsilon^*(\vec{x}) \cr
}\right )' =  \pmatrix{
 \left( { c\lambda (\vec{x})  +d \over c \bar \lambda(\vec{x})   +d }\right)
^{1\over 4}
& 0 \cr
0 &  \left( { c\bar \lambda (\vec{x})  +d \over c  \lambda(\vec{x})   +d
}\right) ^{1\over 4}
 \cr
}
\left (\matrix{
\epsilon (\vec{x})\cr
\cr
\epsilon^*(\vec{x}) \cr
}\right ) \ ,
\end{equation}
\vskip - .3 cm
\noindent where $\lambda (x) =  { {\cal H}_1 \over  {\cal H}_2}
$ and ${\cal H}_{1, 2}(x)$ are  complex harmonic functions defined in eq.
(\ref{harmonic}).
The role of rigid symmetries in connection with the hypermultiplet action in
the black
hole background was discussed before  in \cite{BKO}.

Taking into account the fact that the choice of the gauge condition for the
heterotic string has not required any specific knowledge of available black
holes but only the properties of $N=4$ supergravity under S-duality
(including the spinors!) we may proceed with the duality symmetric gauges for
type II string and supermembrane.

We may choose a scalar matrix in a symmetric gauge
$V=V^{\dagger}$ as suggested in \cite{CJ}. This
condition is provided by the special choice of the local $SU(8)$ gauge-fixing.
\begin{equation}
V =
\pmatrix{
 {1\over \sqrt { 1- Y \bar Y}} &  {Y\over \sqrt { 1- Y \bar Y}} \cr
\cr
 {\bar Y \over \sqrt { 1- Y \bar Y}} &  {1\over \sqrt { 1- Y \bar Y}} \cr
} \ .
\end{equation}

Each entry of this matrix is defined by the matrix  $Y_{AB,CD},
\, A,  \dots , D = 1, \dots ,8$,  i.e. by the  inhomogeneous coordinates of
${E_7
\over SU(8)}$.

 $E_{ 7(+7)}$ acts on the coordinates $Y_{AB,CD}$  by fractional transformation
\begin{equation}
Y' = {B+YD \over A+YC} \ .
\end{equation}
 The action of  U-duality on the scalar matrix $V$ is the following
\begin{equation}
V (Y)   \pmatrix{
A & B \cr
C & D \cr
} = \pmatrix{
U(Y) & 0 \cr
\cr
0 & U(Y)^* \cr
} V(Y') \ ,
\end{equation}
where
\begin{equation}
\pmatrix{
A & B \cr
C & D \cr
}
\end{equation}
is a constant matrix of $E_{ 7(+7)}$, and
 $A, B, C, D $ are 28 by 28  constant matrices defined in \cite{CJ}.
The local $SU(8)$ transformation U is determined by the condition that after
the action of  $E_{ 7(+7)}$ the matrix $V(Y') = V(Y')^\dagger$ remains unitary.

 The local $SU(8)$ compensating transformation which keeps the theory in the
same symmetric gauge after the duality transformation   is defined by
\begin{equation}
\pmatrix{
U(Y) & 0 \cr
\cr
0 & U(Y)^* \cr
} = V (Y)   \pmatrix{
A & B \cr
C & D \cr
} V^{-1} \left ({B+YD \over A+YC}\right ) \ .
\label{compens}\end{equation}
In this form the matrix U still acts on 28-dimensional representation. In the
form in which it acts on a single spinor in an 8-dimensional representation
with $U(Y)_{sp}$ it gives an  explicit transformation of the gauge condition
under U-duality.

Thus we have shown that given a half-supersymmetric background
of $d=11$ or $d=10, N=2$ supersymmetry, the Killing spinor of this
background
 is covariant under the action of U-duality on spinors
(via compensating $SU(8)$).
Therefore if the gauge fixing condition on spinors of type  II superstring or
supermembrane  in a a half-supersymmetric background is chosen in a
manifestly duality covariant form (\ref{zerocharge}),
this gauge fixing will transform into an equivalent one related to the original
one by the
action of  $E_{ 7(+7)}$ (or of some subgroup of it) on gravitino and therefore
on the
Killing spinor.

In absence of the
background we may still use the fact that the algebraic constraint on Killing
spinors at infinity is defined in terms of the $E_{ 7(+7)}$ black hole hair.
The algebraic constraint will transform under duality in terms of the global
$SU(8)$ compensating transformation
\vskip - .3 cm
\begin{equation}
\pmatrix{
U(Y_0) & 0 \cr
\cr
0 & U(Y_{0})^* \cr
} = V (Y_{0})   \pmatrix{
A & B \cr
C & D \cr
} V^{-1} \left ({B+Y_{0} D \over A+Y_{0} C}\right ) \ ,
\label{sym}\end{equation}
\vskip - .3 cm
\noindent where $(Y_{AB,CD})_0$ is the value of the scalar fields
at infinity, or the vacuum  expectation value of the scalar field  $\langle
Y_{AB,CD}\rangle$.
The gauge-fixing condition will transform according to the transformation
property of the spinors.
And again, in symplectic form we have

\begin{equation}
 \left({{\cal Z}\over |Z|} C  \right)' =  \pmatrix{
U(Y_0)_{sp}  & 0 \cr
\cr
0 & U(Y_0)^*_{sp}  \cr
}  \left({ {\cal Z}\over |Z|} C \right) =
0\ , \qquad |Z|' = |Z| \ .
\label{Utransform}\end{equation}
Thus the manifest U-duality is the property of the general
covariant gauge-fixing in terms of the supercharge of the
background and presented in eq. (\ref{zerocharge}). This gauge-fixing
incorporates all possible half-supersymmetric backgrounds of the theory.


\section{Conclusion}

The main result of this work is the gauge-fixing of manifestly supersymmetric
string theory in a form in which duality symmetry is also manifest. The
difference with the known quantization methods is in the fact that string is
placed in the curved BPS background which admits a Killing spinor of the
half-size of the supersymmetry parameter. The gauge-fixing condition uses the
supercharge of the background, see eq. (\ref{zerocharge}), and therefore
general
covariance is not broken in the quantization. The algebraic constraint on the
ghosts
which allows to truncate
$\kappa$-symmetry has a simple form in terms of the symplectic central charge
matrix of the background defined in eq. (\ref{Z}) and is given in eq.
(\ref{sympl}). Under duality transformation of the background this constraint
transforms as shown in eq. (\ref{Stransform}) in doublet form, in eq.
(\ref{Scontracted}) in the contracted form, and in eq. (\ref{Ssympl}) in
symplectic form for
S-duality and in eq. (\ref{Utransform}) in the symplectic form for
U-duality. The partition function on a torus is by construction gauge
independent, and therefore it  is duality symmetric at least at the formal
level,
before anomalies are taken into account. The absence of anomalies of
$\kappa$-symmetry in the heterotic string background  requires the
possibility to perform the embedding of the spin connection with torsion into
the
non-abelian gauge group \cite{Tonin}. Such embedding was already studied for
different black holes and their uplifted versions. It was found
that it does not work equally well for all configurations, depending on whether
the holonomy group is a subgroup of the gauge group of the heterotic string
or not \cite{BKO1,KOexact}.
Thus we expect that the future study of the anomalies of the GS string in
duality
covariant gauges will help us to understand better the quantum aspects of this
theory which were not yet covered in this paper: here we have only shown how to
generalize the light-cone gauge to the most general possible gauges where the
$\kappa$-symmetry is truncated in terms of the central charges of the BPS
backgrounds.

We consider the quantization in the absence of the curved background as a
limiting procedure when curvature goes to zero. This means that we can use  the
central charge of the background even in the flat space to fix the fermionic
symmetry.
The central charge can be considered as
 the vacuum expectation value of the string momenta, see eq.
(\ref{spont}). It was explained in \cite{K95} that the saturation of the
four-dimensional
BPS bound $m_4 = |Z|$ is the condition that the ten-dimensional configuration
is massless, since $ m^2_{10}  =  m^2_{4}  -|Z|^2 =0$ . The massless
ten-dimensional state may correspond also to the massless four-dimensional
state. The form of the gauge-fixing  fermionic symmetry (\ref{Z}), which we are
using has a well defined limit when $|Z|\rightarrow  0$ which allows us to
perform the quantization both in the massless black hole background
(\ref{dilatonmassless}) as well as in the flat  background.

Our main conclusion is the following. The nature of infinite
reducible $\kappa$-symmetry, which is the gauge symmetry of the manifestly
supersymmetric versions of the string theory,   provides a clear request for
the existence  of the geometries associated with the
 states saturating the supersymmetric positivity bound: extreme black holes,
strings, membranes, pp-waves. Those geometries-states play a special role in
gravitational theories. What was   considered previously as a  misfortune of
infinite reducibility  of
$\kappa$-symmetry becomes its enormous advantage: the consistent  truncation
can be performed  in the BPS-type backgrounds with one half of unbroken
supersymmetry, which describe the duality invariant geometries. We have found
duality-symmetric gauges for fixing
$\kappa$-symmetry. In those gauges the elementary excitations of the
supersymmetric string described by the first quantized partition function on
the torus are duality invariant. The second quantization of this theory may
lead to a better understanding of a quantized string and black holes.

\section*{Acknowledgements}

I am grateful to T. Banks, R. Brooks, M. Dine, A. Linde, M. Peskin, J. Schwarz
and  L. Susskind   for  useful discussions.
This  work was  supported by  NSF grant PHY-8612280.


\

\section*{Appendix A: Supersymmetric waves and  black holes}

It was observed some time ago that the uplifted extreme
electrically charged  black
holes become supersymmetric gravitational waves with the Killing spinor
satisfying the light-cone constraint \cite{BKO}.
The most general known to us half-supersymmetric backgrounds of the
heterotic string which admits a Killing spinor satisfying the null condition
$n_\mu
\gamma^\mu
\epsilon =0$ is given by a  bosonic configurations which also admits a
null Killing vector. We call such configurations   supersymmetric
gravitational waves (in the context of string theory they  were called chiral
null
models
\cite{HT}).
Some of them (Brinkmann's plane fronted waves) admit the Killing
vector  which is  covariantly constant \cite{BKO1}.  They  were called
supersymmetric string waves (SSW). The configurations related to the SSW by
T-duality also are supersymmetric and were called generalized
fundamental string solutions or dual waves \cite{BEK}.  The
bosonic part of the ten-dimensional background for the heterotic string
which
is defined by the integrability condition for the truncation of
$\kappa$-symmetry is given by such solutions\footnote{More
general ten-dimensional solutions may still be discovered and at the moment of
this writing there is no information available about the most general
half-supersymmetric backgrounds for the heterotic string. However, on the
basis of the $N=2, d=4$ investigations of Tod \cite{Tod} one may expect that
all
such backgrounds may be listed.}.
Here we would like to consider the half-supersymmetric
$SO(8)$-symmetric  solutions  build from  ten
functions of transverse $x^i,\, i=1, \dots 8$. This  metric  admits
two null Killing vectors $n,m$ , being independent on $u$- and $v$-coordinates.
Indeed, the
gauge-fixing in the generalized $(m,n)$ light-cone gauge in the presence of
the background   naturally requires the background to admit two null
vectors besides a half-size Killing spinor.

The metric is
\begin{equation}
ds^2 =  2e^{2\hat \phi} du (dv  + A_{\mu} dx^\mu )- \sum_1^8dx^i dx^i,
\qquad A_v =0 \ .
\label{wave}\end{equation}
The 2-form field is
\begin{equation}
B = 2 e^{2\hat \phi} du \wedge (dv + A_{\mu} dx^\mu) \ .
\end{equation}
The ten-dimensional dilaton $e^{-2\hat \phi}$ and the $u$-component of the
field $A_\mu$ are harmonic functions  in the eight-dimensional flat space,
\begin{equation}
\sum_1^8\partial_i \partial_i e^{-2\hat \phi} =0 \ , \qquad
\sum_1^8 \partial_i \partial_i A_u =0 \ .
\end{equation}
The eight transverse functions $A_i$ satisfy the following
equations:
\begin{equation}
\sum_1^8 \partial_i (\partial_{[j} A_{i]})=0 \ .
\end{equation}
For some of the  solutions (not all of them) it is
known how to proceed with the spin embedding to cancel
$\alpha'$-corrections coming from anomalies \cite{BKO1,BEK}.

It is instructive to present here also the form of these solutions in terms of
the four-dimensional geometry, when the functions describing the
ten-dimensional waves depend only on $x^1, x^2, x^3$. Dimensional reduction
of supersymmetric gravitational waves was performed in
\cite{KB}. The stationary metric in the canonical frame is
\begin{equation}
ds^2 = e^{2\phi}(dt + \sum_1^3 A_i dx^i)^2 - e^{-2\phi}\sum_1^3 dx^i
dx^i \ ,
\label{bhtype}\end{equation}
where the four-dimensional dilaton is given by
\begin{equation}
e^{-2\phi} = (e^{-2\hat \phi} A_u - \sum_4^8 (A_i )^2 )^{1\over
2} \ .
\label{4dilaton}\end{equation}
The other fields can be also deduced from the ten-dimensional configuration
and are presented explicitly in \cite{KB}. If we would take a special
subclass of dimensionally reduced gravitational waves (\ref{wave}) with
\begin{equation}
A_1=A_2=A_3 =0 \ ,
\end{equation}
we will get the supersymmetric black holes with metric
\begin{equation}
ds^2 = e^{2\phi}dt ^2 - e^{-2\phi}\sum_1^3 dx^i
dx^i \ ,
\label{spher}\end{equation}
where the four-dimensional dilaton is defined in eq. (\ref{4dilaton}). The
functions defining the solutions are taken in the form \cite{KB}:
\begin{equation}
e^{-2\hat \phi}= 1+ \sum_1^s {2\tilde m_k\over r_k}\ , \qquad A_u = 1+
\sum_1^s {2\hat m_k\over r_k}\ , \qquad A_i = \sum_1^s {2 (q_k)_i \over r_k}\ ,
\qquad i=4, \dots 8.
\end{equation}
This is a multi-black-hole solution with S black holes and $r_k \equiv |\vec x-
\vec x_k|$. The four-dimensional dilaton is given by
\begin{equation}
e^{-2\phi}= \left((1+ \sum_1^s {2\tilde m_k\over r_k}) ( 1+
\sum_1^{s} {2\hat m_l\over r_l}) -\left[\sum_1^s {2 (q_k)_i\over r_k}\right]^2
\right)^{1\over
2} \ .
\label{multi}\end{equation}
If we would take one-black-hole solutions with all functions
$A_i=0$ we would reproduce the supersymmetric electrically charged
2-parameter   black hole solutions of Sen \cite{Sen} for $g=1$. The canonical
metric is given in eq. (\ref{spher}), where the dilaton is given by
\begin{equation}\label{DILATON}
e^{-2\phi}= \left(1+ {2(\tilde m +\hat m)  \over r}+ {4 \tilde m \hat m \over
r^2}
\right)^{1\over 2} \ .
\end{equation}
One can rescale this solution by introducing $\langle e^{-2\phi}\rangle  =
e^{-2\phi_0}=
{1\over g^2}$.  This allows to bring our dimensionally reduced wave solution
to the form of eq. (\ref{dilaton}). Two independent parameters in the wave
solutions are related to those in (\ref{dilaton}) as follows:
\begin{equation}\label{MASS}
\tilde m + \hat m = 2 m G_N\ ,  \qquad \tilde m  \hat m = g^2 (N_L - 1) \ .
\end{equation}
The first parameter
is the mass of the black hole, which  in our case is $m= {\tilde m+\hat m
\over
2 G_N}$. The second parameter is  related to the left-handed charge of the
black
hole and  to the parameter $m_0^2  =  4  \tilde m \hat m$ introduced by Sen
\cite{Sen}. Thus if one would wish to generalize Sen'solutions to the
multi-black hole
case using the duality rotations from Kerr's
four-dimensional black holes, this would be very difficult. However, a
duality between ten-dimensional supersymmetric waves and
four-dimensional black holes established in  \cite{BKO} helps to get the
most
general in this class multi-black hole solutions defined in eqs.
(\ref{spher}), (\ref{multi}).

Now we may conclude that the  general electrically charged black
hole-type solutions  (\ref{bhtype})   indeed  form the background in which
the heterotic string in Green-Schwarz form is known to be quantized
consistently. The uplifted geometry admits  the Killing spinor and
 two null vectors required
by our choice of the gauge condition. In dimensionally reduced form this
geometry includes all known electrically charged supersymmetric black
holes of the heterotic string theory saturating the BPS bound.

\section*{Appendix B: Supersymmetric massless multi-black holes}

The massless black hole configuration found by Behrndt \cite{Klaus2}
by dimensional reduction of the
T-self-dual supersymmetric wave solutions is given by
$\tilde m=-
\hat m$, see eqs. (\ref{spher}),  (\ref{DILATON}), (\ref{MASS}).  In this
Appendix we will describe a rather nontrivial space-time structure of this
solution and find a more general set of massless black holes.

Note that for the solutions obtained by Sen \cite{Sen} the   black hole mass
$m$ and
the left-handed charge
$Q_L$ have the following parametrization: $m^2 = {m_0^2 \over 16} \cosh^2
\alpha$ and
$(Q_L)^2 =  {g^2\over  2 }
m_0^2 \sinh^2 \alpha$. Therefore the point where the mass of the black hole is
zero
seems to require also the left-handed charge to vanish, i.e., the solution
becomes trivial.
 However, if one starts with ten-dimensional supersymmetric
gravitational waves (\ref{wave}),  the parameters $\tilde m$ and $\hat m$ are
independent, and there
is no obvious reason not to consider  the configuration $\tilde m +
\hat m = 0$. (The configurations with $\tilde m +
\hat m < 0$, which would correspond to a negative ADM mass, would violate
supersymmetric positivity bound.) Bearing in mind that massless
black  holes are not quite usual solutions of four-dimensional gravity
interacting with
matter,  one may try to study these configurations in more detail.

As we have already mentioned in Sect. \ref{General}, the massless electric
black hole (\ref{dilatonmassless}) has a singularity at $r = 2g$,  in addition
to the singularity ar $r = 0$. Meanwhile, the massless magnetic black hole
(\ref{dilatonmasslessmagn}) has a singularity at $r={2\over g}$. The relevant
question to ask is whether the  singularity of the metric of the electric
black hole at $r=2g$ and of the magnetic one   at  $r={2\over g}$ is a true
singularity, or it can be removed by the change of coordinate system.
To answer this question we calculated the   curvature scalar for these
solutions in canonical Einstein frame. In the electric case the curvature
scalar has a singularity at $r = 2g$:
\begin{equation}
R_{\rm can}^{\rm el} = {4g^2 (2 g^2 + r^2) \over  r (r^2 - 4 g^2)^{5\over 2}} \
{}.
\end{equation}
In the magnetic case the canonical curvature is
\begin{equation}
R_{\rm can}^{\rm magn} = {4g  (2 + g^2 r^2) \over  r (g^2 r^2 - 4 )^{5\over 2}}
\ .
\end{equation}
For completeness of the picture we will check that the new singularity is
present also in stringy frame. For electric solution in stringy frame the
metric is
\begin{equation}
ds^2 =  \left( 1 -  { 4
              g^2 \over r^2} \right)^{-1}  dt^2
 -    d\vec x^2 \ .
\label{dilatonm}\end{equation}
and the curvature is
\begin{equation}
R_{\rm str}^{\rm el} = {-8 g^2 (  8 g^2 + r^2) \over  r ^2 (   r^2 - 4g^2
)^{2}} \ .
\end{equation}
The magnetic massless solutions in stringy frame has the following metric
\begin{equation}
ds^2 =   dt^2
 -   \left( 1 -  { 4
              \over g^2  r^2} \right)d\vec x^2 \ .
\label{dilatonmmagn}\end{equation}
and the curvature scalar  is:
\begin{equation}
R_{\rm str}^{\rm magn} = {16 g^2 (  2+  g^2  r^2) \over   (   r^2 g^2 -4 )^{3}}
\ .
\end{equation}
Thus the singularity at $r=2g$ ($r=2/g)$ does not vanish when we change from
electric to magnetic solutions or change from canonical to stringy frame. For
comparison, we present here the pure magnetic $a=1$ massive black hole in
stringy
frame
\begin{equation}
ds^2 =   dt^2
 -   \left( 1 +  { 4 m
              \over r } \right)d\vec x^2 \ .
\label{dilatonmmagn2}\end{equation}
The curvature is completely non-singular
\begin{equation}
R_{\rm str}^{\rm magn}  (a=1) = {32 m^2  \over   (   4m + r  )^{4}} \ .
\end{equation}
Nothing like that happens with the massless configuration, the singularity
is present and since it is related to string coupling, it somehow reflects the
presence of a string.

One can find a  solution describing a
more general family of four-dimensional black holes with a vanishing
ADM mass. For this purpose we may use the fact that in gravitational wave
solutions
in $d=10$ one can use  more general harmonic functions. For one-black-hole case
one
may take
\begin{equation}
e^{-2\hat \phi}=e^{-2\hat \phi_0} +  {2\tilde m_k\over r}\ ,
\qquad A_u = (A_u)_0 +
 {2\hat m_k\over r}\ , \qquad A_i = (A_i)_0 + {2 (q_k)_i \over r}\ ,
\qquad i=4, \dots 9.
\end{equation}
 The four-dimensional dilaton is now given by
\begin{equation}
e^{-2\phi}= \left((e^{-2\hat \phi_0} +  {2\tilde m_k\over r}) ( (A_u)_0 +
 {2\hat m_k\over r}) -\left[(A_i)_0 + {2 (q_k)_i \over r}\right ]^2
\right)^{1\over
2} \ .
\end{equation}
Obviously there are many ways to make  the ${1\over r}$ term in this
expression equal to zero, and  to make
the ADM mass of the four-dimensional black holes vanishing. The condition on
the
parameters of the harmonic functions which provides the massless black holes is
\begin{equation}
 e^{-2\hat \phi_0} \hat m + (A_u)_0 \tilde  m - 2 \sum_{i=4}^9 (A_i)_0 q_i =0 \
{}.
\end{equation}
The main difference with the previous massless case comes from the
non-vanishing asymptotic value of the  ten-dimensional component of the metric
$g_{ui}=(A_i)_0$. This modification cannot be removed by simple rescaling of
coordinates. Thus generalizing higher-dimensional configurations one can find
more massless four-dimensional black hole-type solutions. Different choices of
harmonic functions in $d=10$ solutions describe different geometries of the
six-dimensional
space. The massless multi-black holes are also available. We may choose the
following harmonic functions in the supersymmetric waves (\ref{wave}).
\begin{equation}
e^{-2\hat \phi}=e^{-2\hat \phi_0} +   \sum_1^s {2\tilde m_k\over r_k}\ ,
\qquad A_u = (A_u)_0 +
\sum_1^s {2\hat m_k\over r_k}\ , \qquad A_i = (A_i)_0 + \sum_1^s {2 (q_k)_i
\over r_k}\ ,
\qquad i=4, \dots 9.
\end{equation}
The total configuration may consists of many massless black holes, the
condition that each black hole in the configuration is massless requires:
\begin{equation}
 e^{-2\hat \phi_0} \hat m_k + (A_u)_0 \tilde  m_k - 2 \sum_{i=4}^9 (A_i)_0 (q_k
)_i=0 \ .
\end{equation}
The existence of massless black hole solutions  presents a new challenge. Some
time ago the very possibility of  black holes being massless would seem
unthinkable.  One could expect that in the limit $m\to 0$ gravitational field
disappears, and space becomes exactly flat.  Now we have a  vector multiplet
which acts a a source of gravity. This leads to existence of  a
 large family of  black holes  which have  nontrivial geometric properties even
  when their ADM mass vanishes. Note that it maybe somewhat misleading   to
call these   states  ``massless black holes.''  First of all, they are massless
in the sense of their ADM mass, however the configuration has a rest frame.
Also,  gravitational attraction becomes increasingly strong near usual black
holes. Meanwhile, in our case massless black holes are in  equilibrium with
each other, and they  gravitationally {\it repel} usual test particles which
come to their vicinity.

 Note, however, that these solutions appear in a situation where we have
massless charged vector fields. In such a situation, just like in the theory of
confinement in QCD, nonperturbative effects may completely change the nature of
charged black hole  solutions. In particular, one may study whether
nonperturbative effects may lead to confinement/condensation of
electrically/magnetically charged massless black holes. Therefore physical
interpretation of  massless black hole solutions and of their possible role in
string theory requires further investigation.

\end{document}